\pdfoutput=1

\documentclass[aip,preprint]{revtex4-2}
\usepackage[utf8]{inputenc}
\usepackage[T1]{fontenc}
\usepackage{graphicx}
\usepackage{amsmath}
\usepackage{multirow}
\usepackage{dcolumn}

\draft

\begin{document}
\title{Phase behaviour of the quantum Lennard-Jones solid}
\author{H. Wiebe$^1$, T. L. Underwood$^2$ and G. J. Ackland$^1$ }

\affiliation{ School of Physics \& Astronomy, The University of Edinburgh, Edinburgh, EH9 3JZ, United Kingdom. } 
\affiliation{ Department of Chemistry, University of Bath, Bath, BA2 7AY, United Kingdom. } 
\begin{abstract}
The Lennard-Jones potential is perhaps one of the most widely-used models for the interaction of uncharged particles, such as noble gas solids. 
The phase diagram of the classical LJ solid is known to exhibit transitions between hcp and fcc phases. However, the phase behaviour of the quantum Lennard-Jones solid remains unknown. 
Thermodynamic integration based on path integral molecular dynamics and lattice dynamics calculations are used to study the phase stability of the hcp and fcc Lennard-Jones solids. 
The hcp phase is shown to be stabilized by quantum effects in PIMD while fcc is shown to be favoured by lattice dynamics, which suggests a possible re-entrant low pressure hcp phase for highly quantum systems.
Implications for the phase stability of noble gas solids are discussed.  
For parameters equating to Helium, the expansion due to zero-point vibrations is associated with quantum melting:  neither crystal structure is stable at zero pressure.
	
\end{abstract}

\maketitle


\section{Introduction}

Since its inception in 1924, the Lennard-Jones (LJ) potential,\cite{Jones1924}
\begin{equation}\label{eq:LJ_potential}
U_{\text{LJ}}(r)=4\varepsilon\biggl[\Bigl(\frac{\sigma}{r}\Bigr)^{12}-\Bigl(\frac{\sigma}{r}\Bigr)^6\biggr],
\end{equation}
has remained the canonical model for short-ranged particle interactions. 
The properties of this potential are uniquely defined due to the presence of only two parameters, $\sigma$ and $\varepsilon$, which set the length and energy scales respectively.  
Despite this simplicity, the LJ system shows remarkably rich phase behavior, including transitions between hexagonal close packed (hcp) and face-centered cubic (fcc) solids.\cite{Kihara1952,jackson2002lattice,Travesset2014,Hertano2016,Calero2016} 
For the classical system hcp is the most stable phase at low temperature and pressure conditions, with fcc becoming preferred upon heating and/or compression. 
However, the free energy differences between hcp and fcc are small, and careful calculation of long-range interactions as well as both harmonic and anharmonic thermal effects are important.\cite{Loach2017,Partay2017}

The attractive $1/r^6$  part of the potential describes van der Waals interactions, making it suitable for studying the noble gas elements. 
Indeed, at low pressures the heavy noble gas elements such as Ar, Xe and Kr\cite{Hansen1969,Rutkai2017}, as well as small molecules like methane,\cite{Saager1990} are well described as classical particles interacting via the LJ potential.
However, as the density of the system is increased or the mass of the particles is decreased, quantum effects become increasingly important.
He, the lightest of the noble gas elements, is dominated by quantum effects.
This has motivated investigations into the \emph{quantum} LJ system, which is far more complicated than the classical system. 
For one thing the uniqueness of the phase diagram is lost: in the quantum system there is an additional free parameter, $\hbar^2/2m$, which corresponds to the "quantumness" of the particles.

The solid phase of the quantum LJ system has been investigated using a number of computational methods.\cite{Herrero2005,Hardy1998,DellaValle1998,Chakravarty2002,Chakravarty2011} 
These include quasi-harmonic lattice dynamics (QHLD), classical molecular dynamics, and path-integral methods. 
QHLD \cite{Fultz2010} is exact in the low temperature limit, and captures quantum effects such as the zero-point energy which are inaccessible to classical methods.
However, it cannot capture anharmonic effects associated with interacting phonons, something well known to play an important role in quantum crystals at moderate temperatures\cite{Cazorla2017}. 
Path integral methods\cite{Ceperley1995,Herrero2014} do not suffer from this shortcoming, capable in principle of providing exact properties of quantum systems. 
These methods exploit the path integral formulation of quantum mechanics, and sample the configuration space of the quantum system using molecular dynamics (PIMD) or Monte Carlo.
However, the downside of these methods is that the computational effort required to obtain accurate results grows exponentially with the quantumness of the system, to the point that highly quantum systems are intractable without resorting to severe approximations.
Thus QHLA and path integral methods are complimentary, the former accurately describing the low temperature limit, and the latter being suitable for moderate temperatures.

The focus of previous studies has been on properties such as the thermal expansion and heat capacity of the quantum LJ solid. 
These studies all assumed fcc to be the stable solid phase. 
However, as mentioned above, the classical LJ solid exhibits regions of both fcc \emph{and} hcp stability, and thus the expectation is that the phase diagram of the quantum LJ solid should also exhibit both these phases. 
Interestingly, an exploration of the solid phase diagram in this system has not yet been undertaken.
Experimentally, while there is some uncertainty with regards to the question of hcp vs. fcc stability in Ne, Ar, Kr  and Xe at low pressures, it appears that the fcc phase is at least metastable in these systems.
\footnote{A review of experimental findings regarding the low pressure phase behaviour of the heavy noble-gas solids can be found in ref.~\onlinecite{Jackson2001}, pp. 159--163.}
By contrast, He is exceptional in that the hcp is the observed phase (notwithstanding a small region of bcc stability near melting at low pressures).\cite{Cazorla2017}
It is not known why He readily takes the hcp structure. 
One might speculate that quantum effects somehow act to stabilise the hcp structure, and that the effect is more pronounced in He than the heavier noble gas elements due to its high quantumness. 
An investigation into the phase behaviour of the quantum LJ solid will shine light on this.

Here, we use PIMD in combination with QHLD to determine the phase diagram of the quantum LJ solid, focusing on low pressure regime. 
We determine the relative stability of hcp and fcc for a range of quantumnesses, ranging from the classical limit to that comparable to He. 
In Section~\ref{sec:methodology} we describe the methodology which underpins our calculations.
Then in Section~\ref{sec:model} we describe our model for the LJ solid, including approximations utilised in our calculations.
Moreover in this section we provide computational details regarding our calculations. 
Results of our PIMD calculations are presented in Section~\ref{sec:pimd_results}, followed by results of our QHLD calculations in Section~\ref{sec:lattice_dynamics_results}. 
In Section~\ref{sec:discussion} we reconcile the PIMD and QHLD results and discuss their implications for the noble gas elements.
Finally, in Section~\ref{sec:conclusion} we restate our main conclusions.

\section{Methodology}\label{sec:methodology}

\subsection{Path integral molecular dynamics}

The path integral formalism exploits the isomorphism between a system of $N$ quantum particles and a set of $\mathcal{P}$ interacting replicas of the system, each consisting of $N$ classical particles. 
The exact quantum partition function $Z=\text{Tr}[e^{-\beta\hat{H}}]$ is mapped onto a classical one\cite{Landau2015} $Z_{\mathcal{P}}$, such that
\begin{equation}\label{eq:Zp}
Z_{\mathcal{P}}=\biggl(\frac{m\mathcal{P}}{2\pi\beta\hbar^2}\biggr)^{3N\mathcal{P}/2}
\int d\mathbf{r}_{1,1}\dotsc\int d\mathbf{r}_{\mathcal{P},N}e^{-\beta H_{\mathcal{P}}}.
\end{equation} 
In this equation $m$ is the mass of the particles, $\beta\equiv 1/(k_BT)$ is the inverse temperature,$\mathbf{r}_{i,j}$ is the position vector of particle $j$ in replica $i$, and

\begin{equation}\label{eq:PI-H}
H_{\mathcal{P}} = \sum_{i=1}^{\mathcal{P}}\sum_{j=1}^N\bigg[ \frac{1}{2}\kappa_{\mathcal{P}}(\mathbf{r}_{i,j}-\mathbf{r}_{(i+1),j})^2 +\frac{1}{\mathcal{P}}\sum_{l=j+1}^NU(|\mathbf{r}_{i,j}-\mathbf{r}_{i,l}|)
\biggr]_{\mathbf{r}_{\mathcal{P}+1}=\mathbf{r}_{1}}
\end{equation}
is the Hamiltonian for the set of replicas.
Note that $Z_{\mathcal{P}}$ is only strictly equal to the quantum partition function $Z$ in the limit $\mathcal{P}\to\infty$. 
However, in practice calculations are necessarily limited to finite $\mathcal{P}$.
Ideally $\mathcal{P}$ is large enough that results are indistinguishable from the $\mathcal{P}\to\infty$ limit.
It can be seen from Eqn.~\ref{eq:PI-H} that each particle interacts with its corresponding particles in adjacent replicas via a harmonic potential with spring constant
\begin{equation}\label{eq:wp}
\kappa_{\mathcal{P}}=\frac{m\mathcal{P}}{\beta^2\hbar^2}=\frac{m\mathcal{P}(k_B T)^2}{\hbar^2}.
\end{equation}
The strength of the inter-replica interactions is therefore directly proportional to both the mass of the particles and the temperature. 
Moreover, the condition $\mathbf{r}_{\mathcal{P}+1}=\mathbf{r}_{1}$ in Eqn.~\ref{eq:PI-H} signifies that $\mathbf{r}_{\mathcal{P}+1,j}=\mathbf{r}_{1,j}$ for all $j$. 
Therefore the replicas form a closed loop, and so the resulting system is often referred to as a ring polymer, with each replica representing a bead in the polymer chain.
From Eqn.~\ref{eq:PI-H} it can also be seen that the particles additionally interact \emph{within} each replica according to the given interatomic potential $U(|\mathbf{r}_{i,j}-\mathbf{r}_{i,l}|)$, which in this case is the LJ potential.

The quantum system described by $H_{\mathcal{P}}$ can be sampled using molecular dynamics techniques. 
To do so, conjugate momenta $\mathbf{p}_{i,j}$ are added to the Hamiltonian such that

\begin{equation}\label{eq:PIMD-H}
H_{\mathcal{P}} = \sum_{i=1}^{\mathcal{P}}\sum_{j=1}^N\bigg[ \frac{\mathbf{p}^2_{i,j}}{2m} +
\frac{1}{2}\kappa_{\mathcal{P}}(\mathbf{r}_{i,j}-\mathbf{r}_{(i+1),j})^2+\frac{1}{\mathcal{P}}\sum_{l=j+1}^NU(|\mathbf{r}_{i,j}-\mathbf{r}_{i,l}|)\biggr]_{\mathbf{r}_{\mathcal{P}+1}=\mathbf{r}_{1}}
\end{equation}

and the extended ring polymer system is evolved in time.
It is important to note these momenta are simply a sampling tool and the resulting dynamics are not representative of the motion of the true quantum system. 
For the quantum LJ solid we are interested only in static properties such as energies, which are calculated over configurational space. 
The presence of the stiff bead-bead harmonic interaction does pose a problem with regards to ergodic sampling, but this issue can be alleviated with aggressive thermostatting techniques such as Nos\'e-Hoover chains\cite{Tuckerman1993} or, more recently, stochastic thermostats\cite{Ceriotti2010} combined with a transformation to normal mode coordinates. 
Further information regarding PIMD algorithms and techniques can be found in refs.~\onlinecite{Marx2009, Herrero2014, Markland2018}.

\subsection{Thermodynamic integration in PIMD}

To evaluate the relative stability of the fcc and hcp LJ solids, we require a comparison of the free energies of both phases. 
We use thermodynamic integration, a robust and widely-used technique for calculating
free energies from MD simulations,\cite{Frenkel2002} to obtain the free energies of both phases.

In thermodynamic integration the free energy difference between two states $\mathcal{A}$ and $\mathcal{B}$ is obtained by introducing a coupling parameter $\lambda$ to the partition function $Z$ of the system such that $Z(\lambda=0)=Z_{\mathcal{A}}$ and $Z(\lambda=1)=Z_{\mathcal{B}}$. 
Since $F=-k_{B}T\ln Z$, the free energy difference can then be accessed as
\begin{equation}\label{eq:thermInt}
\Delta F_{\mathcal{BA}}=F_{\mathcal{B}}-F_{\mathcal{A}}=-\frac{1}{\beta}\int\limits_{0}^{1}\frac{\partial \ln Z}{\partial \lambda} d\lambda.
\end{equation}
If $\mathcal{A}$ is a reference state with a known free energy then the above equation can be used to determine the free energy of state $\mathcal{B}$. 
First $\Delta F_{\mathcal{BA}}$ is determined by integrating $\partial\ln Z/\partial\lambda$ over $\lambda$, and then this $\Delta F_{\mathcal{BA}}$ is added to $F_{\mathcal{A}}$ to obtain $F_{\mathcal{B}}$. 
This approach is routinely used to obtain the free energy of a given classical crystal. 
In this case state $\mathcal{A}$ is chosen to be an Einstein crystal whose free energy is known analytically, while state $\mathcal{B}$ is chosen to be the true crystal. 
The methodology for performing such calculations is well-documented.\cite{Frenkel1984,Vega2008,Aragones2012}

The generalization of thermodynamic integration to a quantum system is straightforward: we must include contributions to the free energy from nuclear quantum effects and so we add a second thermodynamic path between the quantum crystal (state $\mathcal{B}$) and the classical crystal (state $\mathcal{A}$). 
Thus the overall free energy $F$ of the quantum crystal is separated into two terms: the
classical free energy $F_{c}$  plus the excess quantum free energy 
$\Delta F_{q}$:
\begin{equation}\label{eq:dFtot}
F = F_{c}+\Delta F_{q}.
\end{equation}

Nuclear quantum effects in PIMD are controlled by the bead-bead interaction term
\begin{equation}\label{eq:bead-bead}
\bigg(\frac{m\mathcal{P}}{2\beta^{2} \hbar^{2}}\bigg)(\mathbf{r}_{i,j}-\mathbf{r}_{(i+1),j})^2
\end{equation}
and so $\Delta F_{q}$ can be calculated via thermodynamic integration by tuning the strength of this interaction. 
The most straightforward choice is to use the particle mass $\mu$ as the coupling parameter and slowly vary it from the true atomic mass $m_{0}$ to infinite mass in the classical limit:
\begin{equation}\label{eq:dFm}
\Delta F_{q} = \int\limits_{m_{0}}^{\infty}\frac{\partial F(\mu)}{\partial \mu} d\mu =-\frac{1}{\beta}\int\limits_{m_{0}}^{\infty}\frac{\partial \ln Z(\mu)}{\partial \mu} d\mu.
\end{equation}
Evaluation of this derivative yields
\begin{eqnarray}\label{eq:dFprim}
\Delta F_{q} &=& -\int\limits_{m_{0}}^{\infty}\bigg<\frac{3\mathcal{P}}{2\beta\mu}-\sum_{i=1}^{\mathcal{P}}\frac{ \mathcal{P}}{2\beta^{2} \hbar^{2}}(\mathbf{r}_{i,j}-\mathbf{r}_{(i+1),j})^2\bigg>_{\mu} d\mu \nonumber 
\\
&=& -\int\limits_{m_{0}}^{\infty}\frac{\big<T_{\text{prim}}\big>_{\mu}}{\mu} d\mu,
\end{eqnarray}
where $\big<T_{\text{prim}}\big>_{\mu}$ is the primitive estimator for the quantum kinetic energy. 
This estimator has poor convergence properties with large $\mathcal{P}$, however, and so without loss of generality it can be replaced with the more well-behaved centroid-virial estimator $\big<T_{\text{vir}}\big>$,\cite{Vanicek2007,Perez2011,Marsalek2014} where
\begin{equation}\label{eq:Tvir}
\big<T_{\text{vir}}\big>=\bigg<\frac{1}{2\beta}+\frac{1}{2\mathcal{P}}\sum_{i=1}^{\mathcal{P}}\mathbf{r}_{i,j}\cdot\frac{\partial U}{\partial \mathbf{r}_{i,j}}\bigg>.
\end{equation}
Finally, for ease of numerical integration a change of variables to $g=\sqrt{m_{0}/\mu}$ is done\cite{Ceriotti2013,Rossi2015,Fang2016} to allow for integration in the range [0,1]:
\begin{equation}\label{eq:dFtfinal}
\Delta F_{q} = -\int\limits_{0}^{1}\frac{2\big<T_{\text{vir}}\big>_{g}}{g}.
\end{equation}

As this procedure is quite computationally expensive, it is not feasible to use mass thermodynamic integration to fully explore the phase diagram of the quantum LJ solid.
Instead, we chose to perform a full calculation of the quantum-corrected free energy for a single reference point $F_{0}(V_0, T_0)$ and then use Gibbs-Helmholtz integration of the free energy to generate the rest of the phase diagram. 
For a given $(V, T)$ point the free energy can be calculated from the reference point using the thermodynamic relationships
\begin{equation}\label{eq:dFconstV}
    F(T;V) = T \left[ \frac{F_{0}}{T_{0}} - \int\limits_{T_{0}}^{T} \frac{U(T)}{T^2}dT \right]
\end{equation}
for constant volume and 
\begin{equation}\label{eq:dFconstT}
	F(V;T) = F_{0} - \int\limits_{V_{0}}^{V} P(V) dV
\end{equation}
for constant temperature. 
PIMD trajectories must still be run to obtain values for $U(T)$ and $P(V)$, but now only one trajectory is required for each point instead of a full mass thermodynamic integration. 

\subsection{Quasi-harmonic lattice dynamics}\label{sec:harmonic_theory}

QHLD\cite{Fultz2010} entails calculating the phonon density of states for a range of densities, and using these data, in conjunction with equations such as those given below, to calculate physical quantities such as the pressure and free energy.
While this approach is less accurate than PIMD at moderate temperatures, it is insightful. 
In particular, it allows the hcp-fcc free energy difference to be understood in terms of thermal and zero-point vibrational contributions. 
Such a decomposition is not possible in PIMD.

Consider a crystal phase at density $\rho$ and temperature $T$. 
The Helmholtz free energy for the crystal can be decomposed as follows:
\begin{equation}\label{eq:F}
F=U_{\text{GS}}+F_{\text{vib}},
\end{equation}
where $U_{\text{GS}}$ is the ground state energy of the crystal and $F_{\text{vib}}$ is the vibrational contribution to the free energy.
(Here, both $U_{\text{GS}}$ and $F_{\text{vib}}$, are intensive quantities, and similarly for all other energies below). 
In the quasi-harmonic approximation the interatomic forces at the ground states for all $\rho$ are assumed to be harmonic, with $\rho$-dependent force constants. 
In this case $F_{\text{vib}}$ can be expressed as follows:
\begin{equation}\label{eq:Fvibquantum}
  F_{\text{vib}}= F_{\text{zp}}+\frac{k_BT}{N}\sum_{i=1}^{3N-3}\ln\bigg[ 1-e^{-\beta\hbar\omega_i} \bigg],
\end{equation}
where $N$ is the number of particles in the system, $\omega_i$ is the angular frequency of the $i$th phonon (of which there are $3N-3$, excluding the translational modes) for the considered $\rho$,
\begin{equation}\label{eq:Fzp}
    F_{\text{zp}} = \frac{1}{N-1}\sum_{i=1}^{3N-3} \frac{\hbar\omega_i}{2} = \frac{3}{2}\hbar\langle \omega\rangle
\end{equation}
is the zero-point energy, and $\langle \omega\rangle$ denotes the mean phonon frequency. 
In the classical limit, $\hbar\to 0$, it can be shown that
\begin{equation}\label{eq:Fclassical}
  F_{\text{vib}}= \frac{k_BT}{N-1}\sum_{i=1}^{3N-3}\ln\omega_i = 3k_BT\langle\ln\omega\rangle
\end{equation}
(up to an inconsequential temperature-dependent constant).
Another important limit is $T\to 0$. Here the second term in Eqn.~\ref{eq:Fvibquantum} vanishes, leaving $F_{\text{vib}}= F_{\text{zp}}$.

The free energy difference between the hcp and fcc phases at $\rho$, $\Delta F\equiv (F^{\text{hcp}}-F^{\text{fcc}})$ (and similarly for $\Delta U_{\text{GS}}$, $\Delta F_{\text{vib}}$, etc.), can be decomposed similarly to $F$ above.
Using the above equations it can be shown that
\begin{equation}\label{eq:DeltaFquantum}
  \Delta F= \Delta U_{\text{GS}} + \Delta F_{\text{zp}}+\frac{k_BT}{N-1}\sum_{i=1}^{3N-3}\ln\bigg[\frac{ 1-e^{-\beta\hbar\omega_i^{\text{hcp}}}}{1-e^{-\beta\hbar\omega_i^{\text{fcc}}}} \bigg],
\end{equation}
with 
\begin{equation}\label{eq:DeltaFclassical}
  \Delta F= \Delta U_{\text{GS}} + 3k_BT\Delta\langle\ln\omega\rangle
\end{equation}
in the classical limit, and
\begin{equation}\label{eq:DeltaFquantumlimit}
\Delta F =\Delta U_{\text{GS}}+ \Delta F_{\text{zp}}=\Delta U_{\text{GS}}+\frac{3}{2}\hbar\Delta\langle \omega\rangle
\end{equation}
in the zero-temperature limit, where $\omega^{\text{hcp(fcc)}}_i$ denotes the angular frequency of the $i$th phonon for the hcp(fcc) crystal.

Note that, as can be seen from Eqn.~\ref{eq:DeltaFclassical}, $\Delta F$ depends on the hcp and fcc phonon spectra only through $\Delta\langle\omega\rangle$ in the zero-temperature limit. 
Here, quantum vibrational effects favour the structure with lowest zero-point energy, i.e. lowest $\langle\omega\rangle$.
Similarly, it can be seen from Eqn.~\ref{eq:DeltaFclassical} that in the classical limit, vibrational effects act to stabilise the structure with the lowest mean \emph{log-frequency} $\Delta\langle\ln\omega\rangle$. 
Moreover, since the phonon frequencies $\omega_i$ are proportional to $1/\sqrt m$, $\Delta F$ is independent of the mass of the particles in the classical limit because changing $m$ leaves $\Delta\langle\ln\omega\rangle$ unchanged.
By contrast, for the general case (Eqn.~\ref{eq:DeltaFquantum}) $\Delta F$ depends on the masses through $\hbar/\sqrt m$, which is a measure of the quantumness of the system.

The above equations can be used to determine $F^{\text{hcp}}$, $F^{\text{fcc}}$ and $\Delta F$ at a given $\rho$ and $T$ from the hcp and fcc phonon densities of states at $\rho$. 
By considering many $\rho$ and $T$, the regions of the $\rho$--$T$ phase diagram where hcp is stable ($\Delta F<0$) and where fcc is stable ($\Delta F>0)$ can be deduced.
Moreover, it is also possible to use the hcp and fcc densities of states over a range of $\rho$ to determine the $P$--$T$ phase diagram.
This is achieved by first calculating the hcp and fcc Gibbs free energies as functions of $P$ and $T$ via
\begin{equation}
    G(P,T) = F(\rho',T) + P(\rho',T)/\rho,
\end{equation}
where $\rho'$ in this expression is the $\rho$ such that $P(\rho',T)=P$, and
\begin{equation}\label{eq:pressure}
P(\rho,T)=\rho^2\biggl(\frac{\partial F}{\partial\rho}\biggr)_T
\end{equation}
is the pressure at a given $\rho$ and $T$.
Then, the Gibbs free energy difference $\Delta G=(G^{\text{hcp}}-G^{\text{fcc}})$ is evaluated as a function of $P$ and $T$. 
Finally, $\Delta G(P,T)$ is used to deduce the hcp and fcc regions of the $P$--$T$ phase diagram similarly to above for the $\rho$--$T$ phase diagram: $\Delta G<0$ indicates that hcp stable; $\Delta G>0$ that fcc is stable.

\section{Model}\label{sec:model}

\subsection{Reduced units}

We consider a system of $N$ distinguishable quantum particles interacting via the LJ potential (Eqn. \ref{eq:LJ_potential}). As the length and energy scales are set by the $\sigma$ and $\varepsilon$ parameters, it is convenient to define all physical properties of the system in terms of dimensionless reduced units, as in Table \ref{table:units}.
\newcommand\T{\rule{0pt}{2.6ex}}       
\newcommand\B{\rule[-1.2ex]{0pt}{0pt}} 
\begin{table}[!htbp]
	\caption{\label{table:units} Reduced units defined in terms of model parameters $\sigma$ and $\varepsilon$.}
	\begin{ruledtabular}
		\begin{tabular}{cc}
			Quantity & Expression \\
			\hline
			Length & $r^*=r/\sigma$ \T \\
			Energy & $E^*=E/\varepsilon$ \\
			Free energy & $F^* = F/\varepsilon$ \\
			Temperature & $T^*=k_BT/\varepsilon$ \\
			Density & $\rho^*=\rho\sigma^3$ \\
			Pressure & $P^*=P\sigma^3/\varepsilon$ \\
			Time & $t^*=t\sqrt{\varepsilon/(m\sigma^{2})}$ \\
			Quantumness &  $\Lambda^*=\hbar/(\sigma\sqrt{m\varepsilon})$ \\			
		\end{tabular}
	\end{ruledtabular}
\end{table}
The inclusion of quantum effects adds a second lengthscale to the system: the de Boer
parameter\cite{deBoer1957} $\Lambda^*$. 
This is a dimensionless quantity which describes the relationship between the particle diameter $\sigma$ and the de Broglie wavelength of particles with energy $\varepsilon$.
Large values of $\Lambda^*$ indicate a more delocalized quantum system, while $\Lambda^*=0$ corresponds to the classical limit. 
For the noble gases $\Lambda^*$ ranges from $\approx 0.01$ (for Xe) to $\approx 0.4$ (for He).\cite{Cazorla2017} 
Thus we consider the range $\Lambda^*=0$ to 0.4 in this work.

\subsection{Truncation scheme}

In MD simulations it is necessary to truncate the potential interactions in order to avoid artefacts due to self-interaction through the periodic boundary. 
Previous work\cite{Loach2017,Partay2017,jackson2002lattice} has demonstrated that for the classical LJ solid one must take great care with regards to the truncation scheme and the treatment of the long-range interactions. 
The most commonly-used scheme is to shift the potential so that it is continuous at some cutoff radius $r^*_{\text{c}}$:
\begin{equation}\label{eq:shiftedLJ}
	U_{\text{STS}}(r^*)=
	\begin{cases}
		U_{\text{LJ}}(r^*)-U_{\text{LJ}}(r^*_{\text{c}}) & \text{if}\; r^*<r^*_{\text{c}} \\
		0 & \text{if}\; r^*>r^*_{\text{c}}.
	\end{cases}
\end{equation}
This treatment, referred to as the spherically truncated and shifted (STS) model, avoids errors from discontinuous jumps in the potential but it fails to account for the interactions occurring beyond $r^*_{\text{c}}$. 
It therefore displays differing phase behaviour from the "true" LJ potential.\cite{jackson2002lattice} 
Conventional tail corrections,\cite{Frenkel2002} which assume that the radial distribution function $g(r^*)$ is uniform and equal to 1 at $r^*>r^*_c$, are not useful for our purposes because they are independent of the crystal structure.

Instead, the contributions of these long-range interactions to the total energy of the system can be accounted for using what we will refer to here as the ground state perturbation (GSP) model.\cite{jackson2002lattice} 
Here the ground state of the LJ system, where all particles reside on their lattice sites, is treated exactly and only the excitations of the system are subject to truncation. 
To do so, we introduce a correction term to the potential:
\begin{equation}
    U_{\text{GSP}}(r^*)=U_{\text{STS}}(r^*)+U_{\text{LRC}},
\end{equation}
where $U_{\text{LRC}}$ is defined by
\begin{equation}\label{eq:LRC}
	U_{\text{LRC}}=U_{\text{GS}}(\rho^*)-\frac{1}{2}\sum_{i,j:R_{ij}<r_{c}}U_{\text{STS}}(R_{ij}),
\end{equation}
$U_{\text{GS}}(\rho^*)$ is the ground state energy of the untruncated "true" LJ system at density $\rho^*$ and $R_{ij}$ is the inter-particle distance in said ground state. 
The $U_{\text{GS}}(\rho^*)$ term is found from lattice-summation\cite{Jones1925, Kihara1952, Barron1955, Stillinger2001} as
\begin{equation}
U_{\text{GS}}(\rho^*)=2\bigg[\bigg(\frac{\rho^*}{\sqrt{2}}\bigg)^4 A_{12}-\bigg(\frac{\rho^*}{\sqrt{2}}\bigg)^2 A_6\bigg].
\end{equation}
The $A_{12}$ and $A_{6}$ terms have been tabulated for different phases of the LJ solid, and in this work parameters for the fcc and hcp phases were taken from ref. \onlinecite{Jackson2001}. 
For simulations in the $NVT$ ensemble where density is constant, the $U_{\text{LRC}}$ term simply amounts to a constant shift in the relative fcc and hcp energies and will therefore not affect the dynamics of either system. 

\subsection{Computational details}

Classical and PIMD simulations were performed with LAMMPS\cite{Plimpton1995} using the i-PI wrapper.\cite{i-PI} 
Simulations were run with $\sigma=2.96$ \AA{} and $\varepsilon=0.00295$ eV, parameters which have been shown to give good results for PIMC calculations of noble gas solids.\cite{Chakravarty2002,Chakravarty2011} 
Systems of 256 Lennard-Jones particles with either the fcc or hcp crystal structures were initialized at a specified density $\rho^*$. 
Trajectories were then initiated in the \textit{NVT} ensemble with orthorhombic periodic boundary conditions.
Temperature was kept constant using the stochastic PILE-G thermostat\cite{Ceriotti2010} with a relaxation time of 0.01 $t^{*}$. 
Simulations were run at temperatures ranging from $T^*=$ 0.10 to 0.50 and $\rho^*=$ 0.65 to 1.30, well below the melting curve for the classical LJ solid.\cite{Morris2002,Mastny2007} 
A timestep of 0.001 $t^{*}$ was used for both classical MD and PIMD. 
For all PIMD phase diagrams the reference point was chosen to be at $T^*_0 = 0.10$ and $\rho^*_0 = 1.07255$, which is the zero-pressure density for the classical LJ solid.

For our QHLD calculations we used the code GULP \cite{Gale2003} to calculate phonon density of states (DoS); and to calculate static crystal energies we used the lattice-sum-based expressions provided in ref.~\onlinecite{jackson2002lattice}.
In the GULP calculations we employed a 12-atom orthorhombic unit cell for both hcp and fcc. 
The hcp unit cell corresponded to six planes stacked in the $z$-direction, with two particles per plane, and a stacking sequence of ABABAB. 
The fcc unit cell was the same except the stacking sequence was ABCABC, this ensures both structures have the same reduced Brillouin zone. 
The accuracy and precision of the DoS output by GULP is determined by how many k-points, $N_{\text{k-points}}$, are used in sampling the Brillouin zone, and how many bins $N_{\text{bins}}$ are used in the DoS histogram -- which is the key output by GULP for our purposes. 
We performed preliminary calculations at $\rho^*=0.8, 1.07255$ and 1.3 in order to determine appropriate values for these parameters, and found that a Monkhorst-Pack scheme with 40 grid points along each dimension of the Brillouin zone, and 300,000 bins in the DoS histogram, was sufficient.
To elaborate, these parameters yielded an error for $\Delta\langle\omega\rangle$ (associated with numerical integration over the DoS histogram) which was significantly less than $|\Delta\langle\omega\rangle|$, ensuring that the calculations had sufficient precision to distinguish which of hcp and fcc was stable according to the zero-point energy (see Section~\ref{sec:harmonic_theory}). 
These parameters also yielded values for $\Delta\langle\omega\rangle$ which were converged with respect to both $N_{\text{k-points}}$ and $N_{\text{bins}}$.
Finally, we note that we found it necessary to modify GULP's source code for this work. Specifically, we increased the number of significant figures used in the output file containing the DoS histogram; the default output format lacked the precision to reduce the error in $\Delta\langle\omega\rangle$ to what was required for this work.

\section{PIMD Results}\label{sec:pimd_results}

\subsection{Classical free energies}

As a starting point for the PIMD calculations, we first determined the classical free energy term $F^*_{c}$ at the chosen reference point of $T^*_0 = 0.10$ and $\rho^*_0 = 1.07255$ using the standard Einstein crystal method\cite{Frenkel1984,Vega2008,Aragones2012}. 
The interaction cutoff $r_c^*$ was initially chosen to be $2.5$, as it is the most popular choice in the literature. 
The sensitivity of the $F^*_{c}$ term to the choice of cutoff length $r^*_c$ was investigated by running simulations with $r^*_c=2.2$ and $2.8$ as well. 
The results are listed in Table \ref{table:dFc}, with the phase stability expressed relative to the hcp phase as
\begin{equation}\label{eq:dF}
\Delta F^*_c = F^*_{c,\text{hcp}} - F^*_{c,\text{fcc}}.
\end{equation}

\begin{table*}[!htbp]
	\caption{\label{table:dFc} Classical free energy differences for the fcc and hcp phases in units of $\varepsilon$ per atom, obtained using thermodynamic integration at $T^* = 0.10$ and $\rho^*_0 = 1.07255$. Negative values of $\Delta F^*_{c}$ indicate stability of the hcp phase. Results are shown for the spherically truncated and shifted (STS) model as well as the correction for the ground state perturbation (GSP model). Errors were obtained using block averaging.}
	\begin{ruledtabular}
		\begin{tabular}{cddddd}
			\multirow{2}{*}{\boldmath{$r^*_c$}} & \multicolumn{3}{c}{\bfseries{STS}} & \multicolumn{2}{c}{\bfseries{GSP}} \\ 
			\cline{2-6}
			& \multicolumn{1}{c}{$F^*_{c,\text{fcc}}$} & \multicolumn{1}{c}{$F^*_{c,\text{hcp}}$} & \multicolumn{1}{c}{$\Delta F^*_{c}$} & \multicolumn{1}{c}{$\Delta U(\rho^*)_{\text{LRC}}$} & \multicolumn{1}{c}{$\Delta F^*_{c,\text{corr}}$} \T \B \\
			\hline \hline
			2.2 & -5.33423(4) & -5.33667(4) & -0.00244(6) & 0.00152 & -0.00092(6)  \T \\
			2.5 & -5.86319(4) & -5.87210(4) & -0.00892(6) & 0.00797 & -0.00094(6) \T \\
			2.8 & -6.18448(4) & -6.17927(4) & 0.00521(6)  & -0.00618 & -0.00097(6)  \T \\
		\end{tabular}
	\end{ruledtabular}
\end{table*}

Predictably, the phase stability of the STS model is highly dependent on the chosen cutoff,\cite{Loach2017, Partay2017} with $r^*_c=2.2$ and $2.5$ favouring hcp while $2.8$ favours fcc. 
To reduce this cutoff effect the GSP correction to $\Delta F^*_c$ was calculated from the ground state fcc and hcp structures using Eqn. \ref{eq:LRC} as $\Delta U(\rho^*)_{\text{LRC}}=U_{\text{LRC}}^{\text{hcp}}(\rho^*_{\text{hcp}})-U_{\text{LRC}}^{\text{fcc}}(\rho^*_{\text{fcc}})$. 
The resulting correction term and corrected free energy difference $\Delta F^*_{c,\text{corr}}=\Delta F^*_c + \Delta U(\rho^*)_{LRC}$ is also listed in Table \ref{table:dFc}. The net result is that the effect of the cutoff length is almost entirely eliminated in the GSP model, with all systems showing the same $\Delta F^*_{c,\text{corr}}$ within error and stabilization of hcp in all cases.

From here, the rest of the phase diagram up to $T^* = 0.5$ and $\rho^* = 1.30$ was calculated using Eqns. \ref{eq:dFconstV} and \ref{eq:dFconstT}. For systems with different densities it was necessary to scale the cutoff length so that the same number of neighbour shells were included inside the cutoff for all simulations. This was done relative to the reference point at $\rho^*_0 = 1.07255$ as
\begin{equation}\label{eq:cutoffScaling}
r_c^*(\rho^*) = 2.5 \left(\frac{1.07255}{\rho^*}\right)^{1/3}.
\end{equation}

\begin{figure} [!htbp]
	\includegraphics[width=0.7\columnwidth]{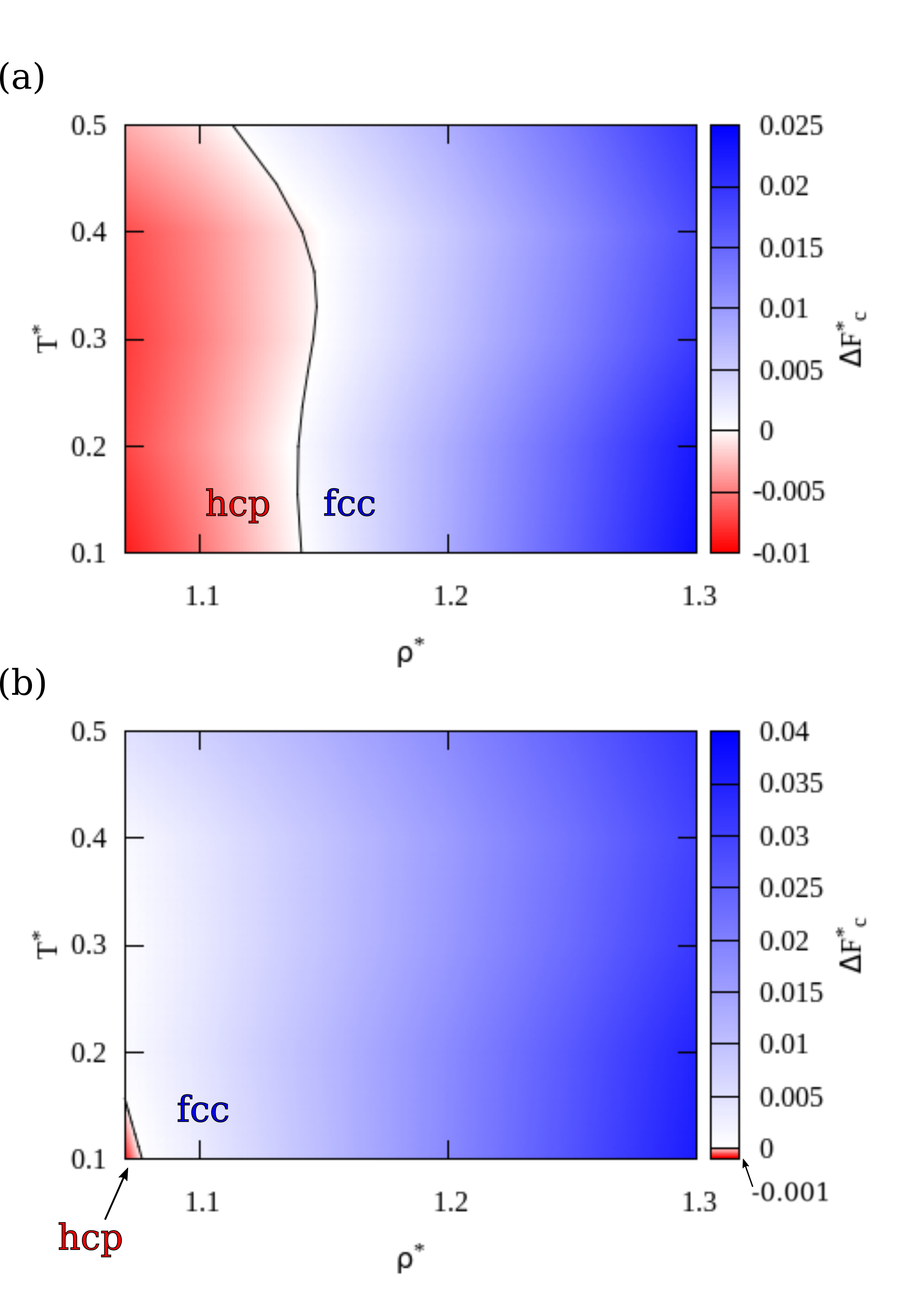}
	\caption{\label{fig:classicalPhaseDiagram} Phase diagrams for the classical LJ solid obtained using thermodynamic integration for (a) the STS model and (b) the GSP-corrected model. Red regions indicate hcp stability, and blue corresponds to fcc.}
\end{figure}

The resulting phase diagrams are shown in Figure \ref{fig:classicalPhaseDiagram} for both the STS and GSP models. 
The STS model shows that hcp is stable at low densities and is in agreement with previous assessments of the phase behaviour of the uncorrected $r_c^* = 2.5$ LJ solid.\cite{Jackson2001} 
Inclusion of long-range GSP correction shifts the phase boundary by stabilizing fcc over hcp, and as such the hcp phase is only stable at low density and temperature. 
Since the fcc phase has been shown to have higher entropy than the hcp phase,\cite{Esbjorn1973,Bolhuis1997,Mau1999,Bruce2000,Elser2014} it is then perhaps unsurprising that inclusion of long-range order works to stabilize fcc. 
All in all, these results demonstrate the delicate balance between solid phases and the surprising complexity in this simple model. 
Further information about the phase stability may be gleaned from a consideration of finite size effects, but due to the large computational cost for PIMD simulations a study of large systems is beyond the scope of this work.

\subsection{PIMD convergence}
\begin{figure}[!htbp]
	\includegraphics[width=0.7\columnwidth]{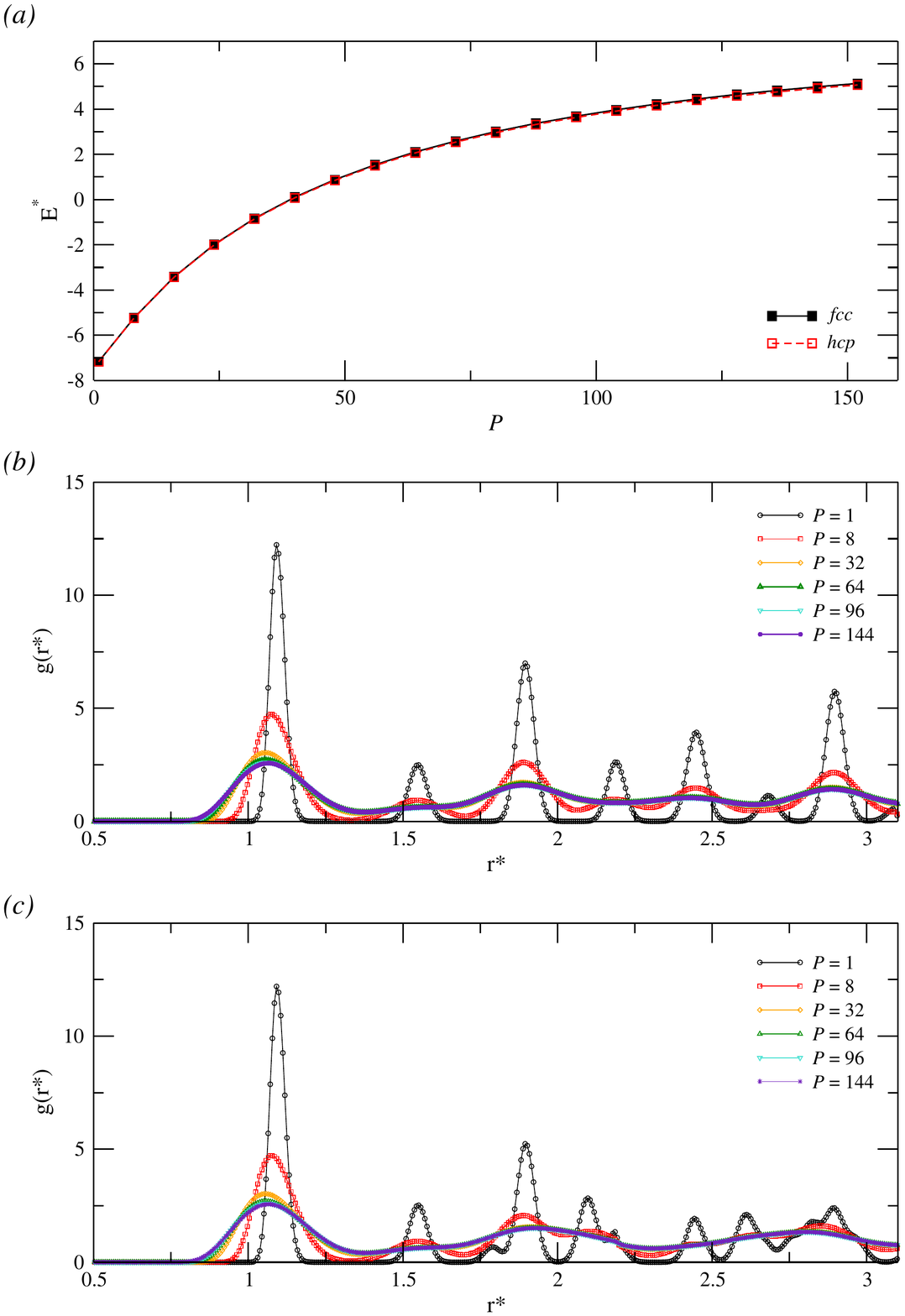}
	\caption{\label{fig:PIMD convergence} Convergence of PIMD simulations with respect to $\mathcal{P}$. \textit{(a)} $\big<E^*\big>$ with $\mathcal{P}$ for fcc (black circles) and hcp (red squares) at $T^{*}=0.1$ and $m=1$. Statistical sampling errors are smaller than the line thickness. Radial distributions function of the \textit{(b)} \textit{fcc } and \textit{(c)} \textit{hcp } phases for $\mathcal{P}=$ 1 to 144 at the same conditions, showing structural convergence.} 
\end{figure}

The number of beads used in a PIMD simulation is a very important choice; $\mathcal{P}$ must be large enough to accurately probe the quantum limit, but small enough that the computational cost is still affordable. 
The number of beads required depends on the relative strength of the quantum harmonic energy levels versus the thermal energy. 
A typical rule of thumb given for the minimum number of beads is $\mathcal{P}_{min}=4\beta\hbar\omega_{\text{max}}$,\cite{Herrero2014} where $\omega_{\text{max}}$ is the highest vibrational frequency in the system. 
However, in practice the required number of beads is often much higher than this minimum limit. 
To choose an appropriate number for the quantum LJ solid, PIMD simulations were run with increasing $\mathcal{P}$ and the system was deemed to be converged when the average internal energy $\big<E^*\big>=\big<U^*_{LJ}\big>+\big<T^*_{\text{vir}}\big>$ was within 0.34 $\varepsilon$, which corresponds to $\approx$ 1 meV/atom in real units. 
The convergence of the structural properties of the system was also monitored via the radial distribution function of the ring polymer beads. 
Convergence was reached at $\mathcal{P}=144$ for $T^{*}=0.10$. Convergence plots are shown in Figure \ref{fig:PIMD convergence} for both energetic and structural properties. 

\subsection{Quantum free energies}

\begin{figure}[!htbp]
	\includegraphics[width=0.7\columnwidth]{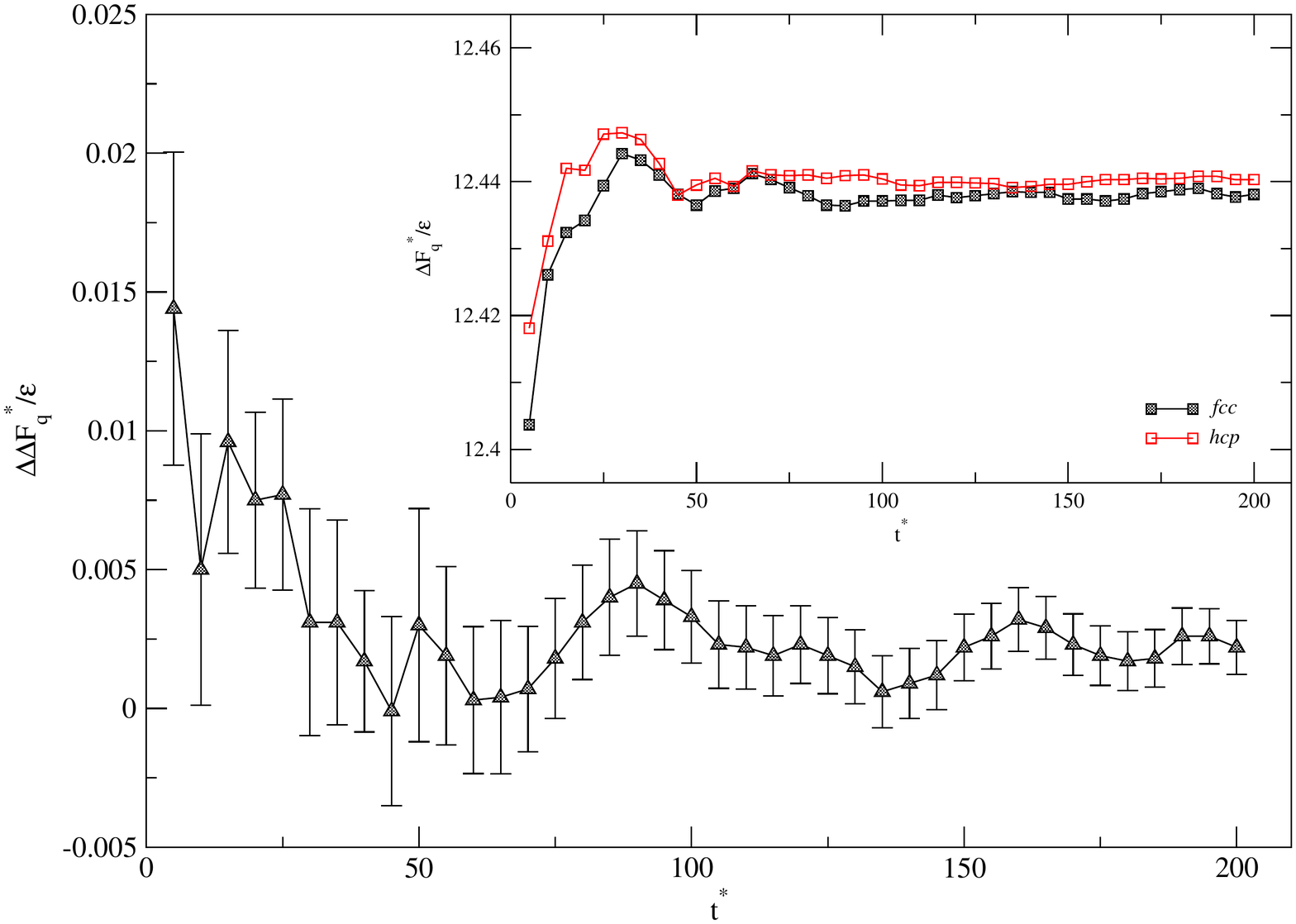}
	\caption{\label{fig:PIMD thermoInt convergence} Convergence of $\Delta\Delta F^*_{q}$ with respect to PIMD trajectory length for $T^{*}=0.10$ and $m_{0}=1$. An initial equilibration period of 5 $t^*$ is not included in this plot. The inset shows the convergence of the individual fcc and hcp $\Delta F^*_{q}$ values. }
\end{figure}

The excess quantum free energies $\Delta F^*_{q}$ for each phase were obtained from PIMD using a 13-point mass thermodynamic integration from $m=m_0$ to $m=\infty$ at the chosen reference point $T^*_0 = 0.10$, $\rho^*_0 = 1.07255$. The free energy difference 
\begin{equation}
\Delta\Delta F^*_{q}=\Delta F^*_{q,\text{hcp}}-\Delta F^*_{q,\text{fcc}}
\end{equation}
between the two phases was monitored and trajectories 150-200 $t^{*}$ in length were required for adequate convergence of this quantity. An example is shown in Figure \ref{fig:PIMD thermoInt convergence}. 
Five $m_0$ values were chosen such that the solid ranged in quantumness from $\Lambda^* =$  0.1 (20 amu) to 0.4 (1 amu). 

\begin{figure*}[!htbp]
    \centering
    \includegraphics[width=\textwidth]{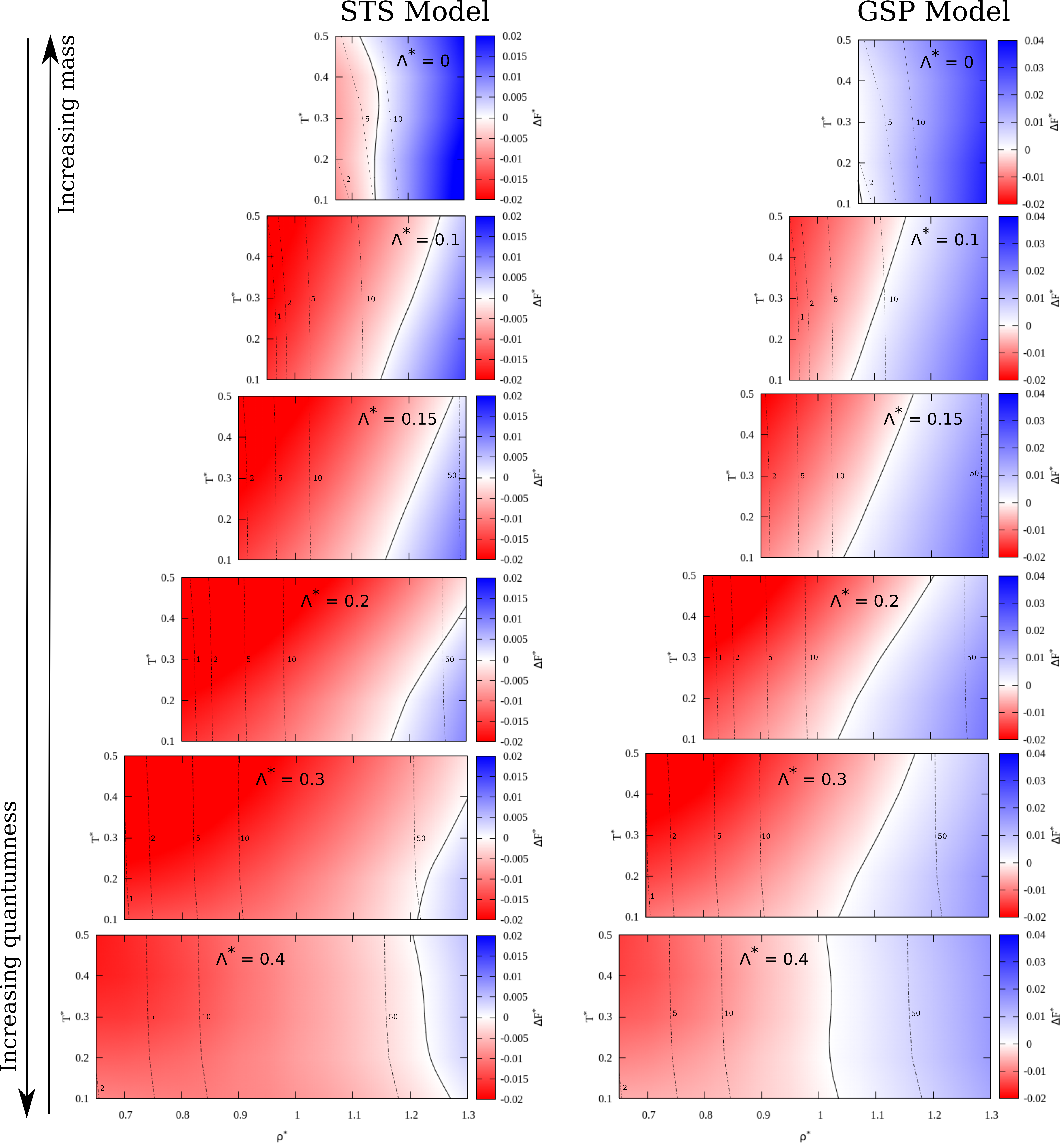}
    \caption{Phase diagrams for the quantum LJ solid. As in Figure \ref{fig:classicalPhaseDiagram}, red regions correspond to hcp stability and blue correspond to fcc stability. The STS model is shown on the left, and the GSP model on the right. The columns are arranged in order of increasing quantumness, with $\Lambda^* = 0$ at the top and $\Lambda^* = 0.4$ at the bottom. Dashed lines correspond to isobars for the hcp phase at $P^* =$ 1, 2, 5, 10 \& 50. Isobars for the fcc phase are indistinguishable from the hcp phase on the scale of this plot.}
    \label{fig:quantumPhaseDiagrams}
\end{figure*}

The phase stability of the quantum LJ solid was then calculated using Eqns.~\ref{eq:dFconstV} and \ref{eq:dFconstT} with $F^*_{0} = F^*_{c} + \Delta F^*_{q}$ at the reference point of $T^*_0 = 0.10$, $\rho^*_0 = 1.07255$. 
For each $\Lambda^*$ value, calculations were run from $\rho^* = 1.30$ to whichever density gave $P^*\approx0$. 
Note that only densities where the system remained solid across the whole temperature range were considered here, thus in some cases it was not possible to reach $P^*\approx0$ due to melting at higher temperature. The resulting phase diagrams are shown in Figure~\ref{fig:quantumPhaseDiagrams} for both the STS and GSP models. 
Phase stability is represented as $\Delta F^* = F^*_{\text{hcp}} - F^*_{\text{fcc}}$.

The quantum contribution to the pressure is immediately apparent in these plots, since the highly quantum systems require significantly lower densities to reach $P^*\approx0$ than in the classical case. 
Even though the GSP model favours fcc more than the STS model, the inclusion of nuclear quantum effects stabilizes hcp over fcc in both cases. 
This is evidenced by the increase in size of the hcp region with increasing quantumness, which is observed even with a relatively heavy particle at $\Lambda^* = 0.1$. 
Since these phase diagrams are plotted against $\rho^*$, this growth of the hcp region means that the fcc region is being pushed to higher and higher pressures with increasing quantumness and is therefore being destabilized relative to hcp. 
Interestingly, the slope of the phase boundary changes sign at $\Lambda^* = 0.4$. 
This may be due to fluctuations in this highly quantum system, or it may indicate a change in the nature of the phase boundary at high quantumness.

\section{QHLD Results}\label{sec:lattice_dynamics_results}

\subsection{Initial investigations}

\begin{figure}[!htbp]
	\includegraphics[width=0.7\columnwidth]{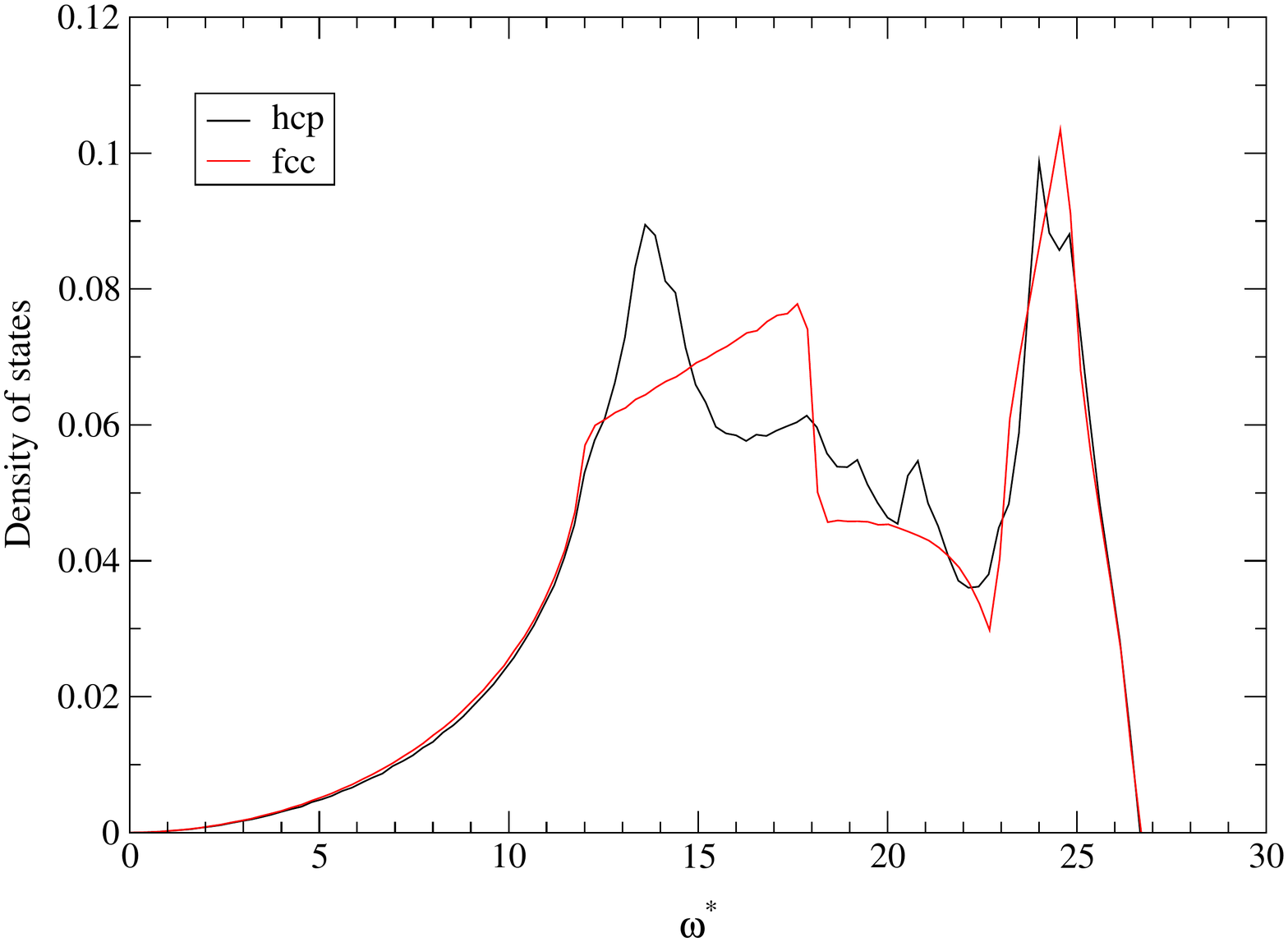}
	\caption{Phonon density of state for the hcp and fcc structures at $\rho^*=1.07255$ for the LJ solid with $r_c^*=10.0$.}
	\label{fig:dos}
\end{figure}

The hcp and fcc phonon density of states (DoS) for $r_c^*=10.0$ at $\rho^*=1.07255$, the density for the classical LJ solid at $T^*=P^*=0$, are shown in Figure~\ref{fig:dos}.
The figure shows that the hcp DoS has both low and high frequency peaks, while fcc has more intermediate frequency modes.
Moreover, both structures share the same peak at high frequency, though the high-frequency peak is larger in fcc than hcp.
However, despite the qualitatively different shapes of the densities of states of the two crystals, the mean frequencies and mean log-frequencies, which, as discussed in Section~\ref{sec:harmonic_theory}, play an important role in determining which of the structures is stable, are indistinguishable on the scale of this figure. 
Hence the fine structure of the DoS must be considered to determine which of hcp and fcc is stable. 
We return to this point in a moment.

Preliminary calculations revealed that the hcp and fcc crystals were mechanically unstable for
densities less than $\rho^*=0.8$: at these densities the system exhibited phonons with imaginary frequencies for all cutoffs considered. 
Hence, keeping in mind that we are interested in the low pressure region of the phase diagram, we focused on densities ranging from $\rho^*=0.8$ to 1.3.

\begin{figure}[!htbp]
    \centering
    \includegraphics[width=0.7\columnwidth]{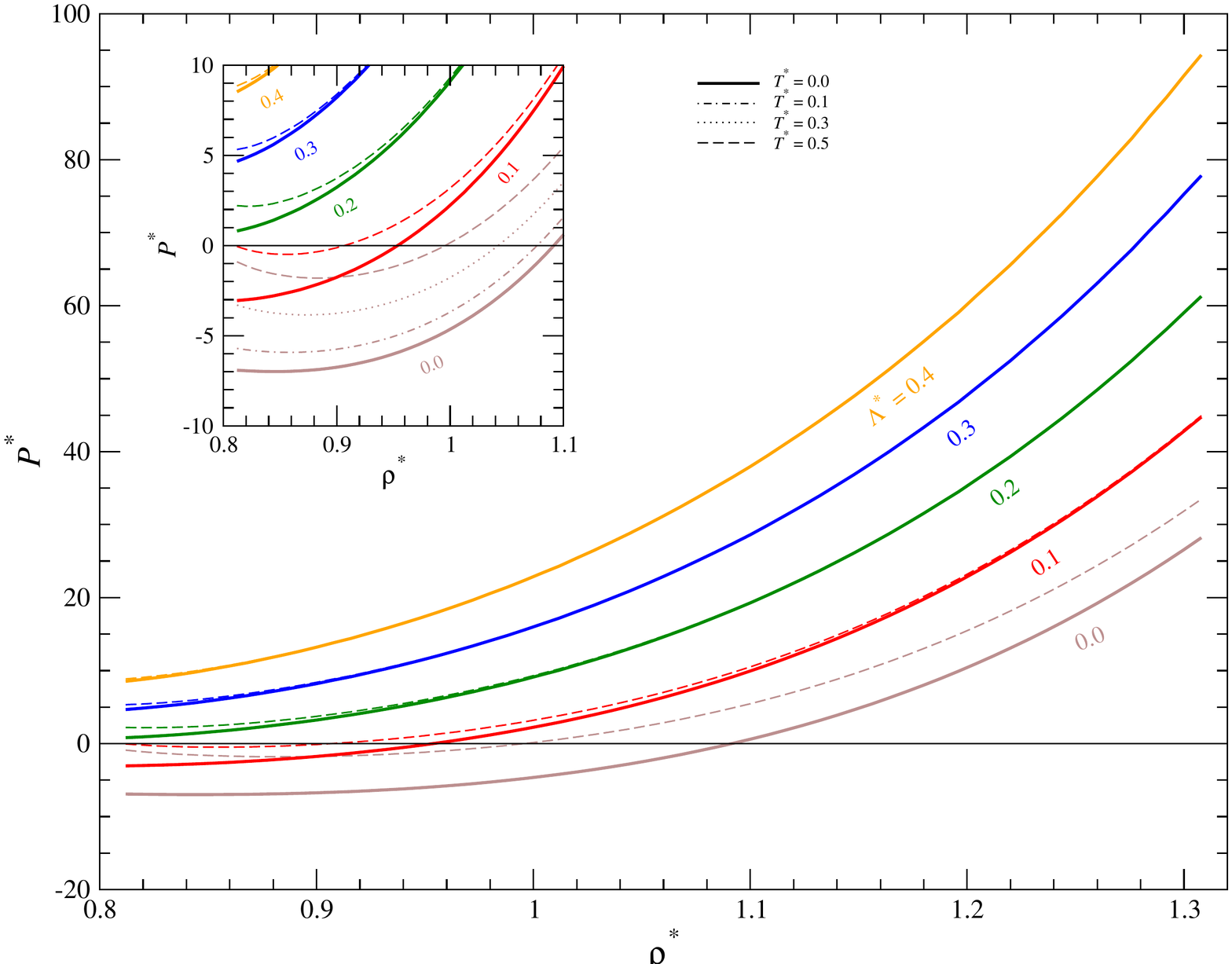}
    \caption{$P^*$ vs. $\rho^*$ for the hcp phase of the LJ solid ($r_c^*=10$) obtained from QHLD for various $\Lambda^*$ 
    and $T^*$. The inset shows the low density region in more detail.}
    \label{fig:harmonic_pressure}
\end{figure}

The hcp pressure is shown as a function of $T^*$ over this density range for various $\Lambda^*$ in Figure~\ref{fig:harmonic_pressure}. 
The fcc pressure is indistinguishable from that of hcp on the scale of this figure. 
As expected, increasing the quantumness while fixing the density results in an 
increase in the pressure of the system. 
This is primarily due the zero-point vibrations. 
To elaborate, from Eqns.~\ref{eq:Fzp} and \ref{eq:pressure} the contribution to the pressure from this energy is
\begin{equation}
P_{\text{zp}}=\rho^2\biggl(\frac{\partial F_{\text{zp}}}{\partial\rho}\biggr)_T=\frac{3}{2}\rho^2\frac{\partial \langle\hbar\omega\rangle}{\partial\rho}.
\end{equation}
Noting that $\partial\langle\hbar\omega\rangle/\partial\rho$ is positive and proportional to $\hbar/\sqrt{m}$, and hence also $\Lambda^*$, it follows that $P_{\text{zp}}$ is also proportional to $\Lambda^*$.
Figure~\ref{fig:harmonic_pressure} also reveals that for $\Lambda^*=0.2$, 0.3 and 0.4 there is no \emph{mechanically stable} hcp or fcc density corresponding to $P=0$ in the quasi-harmonic approximation: the figure implies that the hcp and fcc densities for $P=0$ would be achieved at $\rho<0.8$, which, as just mentioned, are mechanically unstable within the approximation. 
Since the quasi-harmonic approximation is valid in the low temperature limit, for $\Lambda^*\geq 0.2$ some phase other than hcp or fcc must therefore be stable at $P=T=0$.

To validate our implementation of QHLD we also considered $\Lambda^*=0.0103$, 0.0166, 
0.0296 and 0.0896, which correspond to Xe, Kr, Ar and Ne respectively,\cite{DellaValle1998} and compared results for $r_c^*=10$ to those of ref.~\onlinecite{DellaValle1998}. 
Plots of $P^*$ vs. $\rho^*$ for these $\Lambda^*$ at $T^*=0$ (not shown) were in good agreement with those given in ref.~\onlinecite{DellaValle1998}, as were plots of $\rho^*$ vs. $T^*$ at $P^*=0$ (not shown).

\subsection{Sensitivity to cutoff}

To investigate the sensitivity of the phase behaviour to $r_c^*$, we focused on $\rho^*=1.07255$. 
We considered $r_c^*=2.2$, 2.5 and 2.8, supplementing our calculations described above for this density using $r_c^*=10$. 
The hcp and fcc DoS for $r_c^*=2.2$, 2.5 and 2.8 are almost identical to those shown in Figure~\ref{fig:dos} (which recall are for $r_c^*=10$) on the scale of the figure. 
However, differences in the fine structure of the DoS for different $r_c^*$ have important implications for the stability of hcp vs. fcc.

Recall that in the zero-temperature limit $\Delta F$ depends on the zero-point energy through the difference in the mean phonon frequencies $\Delta\langle\omega\rangle\equiv(\langle\omega\rangle_{\text{hcp}}-\langle\omega\rangle_{\text{fcc}})$ (Eqn.~\ref{eq:DeltaFquantumlimit}), while in the classical limit $\Delta F$ depends on the difference in the mean log-frequencies $\Delta\langle\ln\omega\rangle$ (Eqn.~\ref{eq:DeltaFclassical}).
In Table~\ref{table:meanfreqs} $\Delta\langle\omega^*\rangle$ and $\Delta\langle\ln\omega^*\rangle$ are compared for various values of $r_c^*$. 
It can be seen that $\Delta\langle \ln \omega^*\rangle$ is positive for all $r_c^*$. 
This implies that vibrational effects act to stabilise fcc in the classical limit for all considered $r_c^*$. 
This follows from Eqn.~\ref{eq:DeltaFclassical}: if $\Delta\langle \ln \omega^*\rangle >0$, $\Delta F^*$ increases with $T$, which corresponds to fcc stabilisation.

By contrast, the sign of $\Delta\langle \omega^*\rangle$ depends on $r_c^*$. 
$\Delta\langle \omega^*\rangle$ is negative at $r_c^*=2.2$, implying that the zero-point energy of hcp is lower than that of fcc.
Since increasing quantumness increases the size of the zero-point contribution to $\Delta F^*$ (c.f. Eqn.~\ref{eq:DeltaFclassical}), this means that increasing quantumness stabilises hcp for $r_c^*=2.2$ (in the zero-temperature limit).
On the other hand, for $r_c^*=2.5$, $\Delta\langle \omega^*\rangle$ is positive, which implies the opposite, i.e. that fcc is stabilised by quantum effects. 
The same is true for $r_c^*=2.8$ and $r_c^*=10.0$, though for these cutoffs $\Delta\langle \omega^*\rangle$ is of a smaller magnitude and hence the stabilisation of fcc by quantum effects is less pronounced than at $r_c^*=2.5$.

\begin{table}[!htbp]
	\caption{\label{table:meanfreqs} Differences in the phonon mean frequencies and mean log-frequencies between the hcp and fcc	structures,	i.e. $\Delta\langle \omega^*\rangle$ and $\Delta\langle \ln \omega^*\rangle$, for the LJ solid at $\rho^*=1.07255$ and various cutoffs $r_c^*$. A positive value indicates that $\Delta\langle \omega^*\rangle$ or $\Delta\langle \ln \omega^*\rangle$ is higher for the hcp structure than for the fcc structure.}
	\begin{ruledtabular}
	\begin{tabular}{ccc}
	     $r_c^*$ &  $\Delta\langle \omega^*\rangle$ & $\Delta\langle \ln \omega^*\rangle$ \\ 
	     \hline
	     2.2   & -0.0012            & 0.0005                                \\  
	     2.5   &  0.0102            & 0.0013                                \\  
	     2.8   &  0.0021            & 0.0008                                \\  
	     10.0  &  0.0033            & 0.0009                                \\ 
	\end{tabular}
	\end{ruledtabular}
\end{table}

\begin{figure}[!htbp]
	\includegraphics[width=0.7\columnwidth]{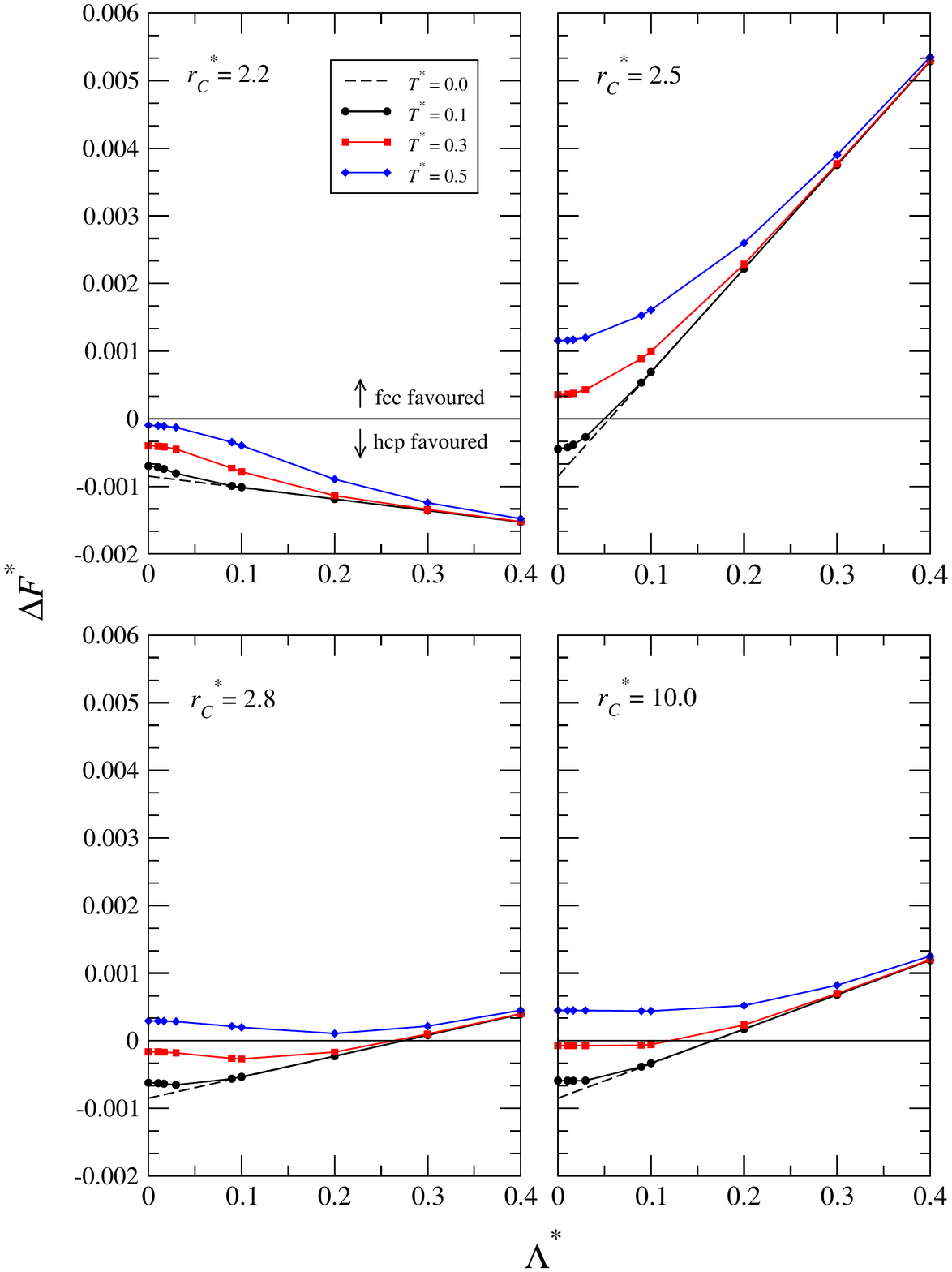}
	\caption{hcp-fcc free energy difference, $\Delta F$, from QHLD vs. de Boer parameter, $\Lambda^*$, for the LJ solid at 
	$\rho^*=1.07255$. Each panel corresponds to a different cutoff $r_c^*$.}
	\label{fig:FvsLambda_harm}
\end{figure}

The above discussion is borne out in Figure~\ref{fig:FvsLambda_harm}, which  shows $\Delta F^*$ vs. $\Lambda^*$ at this density for various $T^*$ and $r_c^*$. 
Note that for $T^*=0$ increasing the quantumness increases $\Delta F^*$ for $r_c^*=2.5$, 2.8 and 10.0, with a more pronounced increase for $r_c^*=2.5$; while increasing quantumness decreases $\Delta F^*$ for $r_c^*=2.2$. 
These trends are in accordance with the values of $\Delta\langle \omega^*\rangle$ for each cutoff described above. 
Note also that, at $\Lambda^*=0$, increasing $T^*$ increases $\Delta F^*$, in
accordance with the above discussion that $\Delta\langle \ln \omega^*\rangle$ is positive for all $r_c^*$, i.e. that vibrational effects always stabilise fcc in the classical limit. 
This is also the case away from the classical limit: \emph{thermal effects} always act to stabilise fcc.

A key result to be drawn from the above discussion is that the difference in the zero-point energies of the hcp and fcc structures in the LJ solid is highly sensitive to $r_c^*$, to the point that changing $r_c^*$ can reverse whether hcp or fcc is stablised by increasing the quantumness of the system. 
There is no analogous problem in the classical case; the discussion above, and previous studies, have revealed that changing $r_c^*$ does not change the fact that the fcc structure is stabilised by thermal effects.

\subsection{Phase behaviour}

\begin{figure}[!htbp]
    \centering
    \includegraphics[width=0.7\columnwidth]{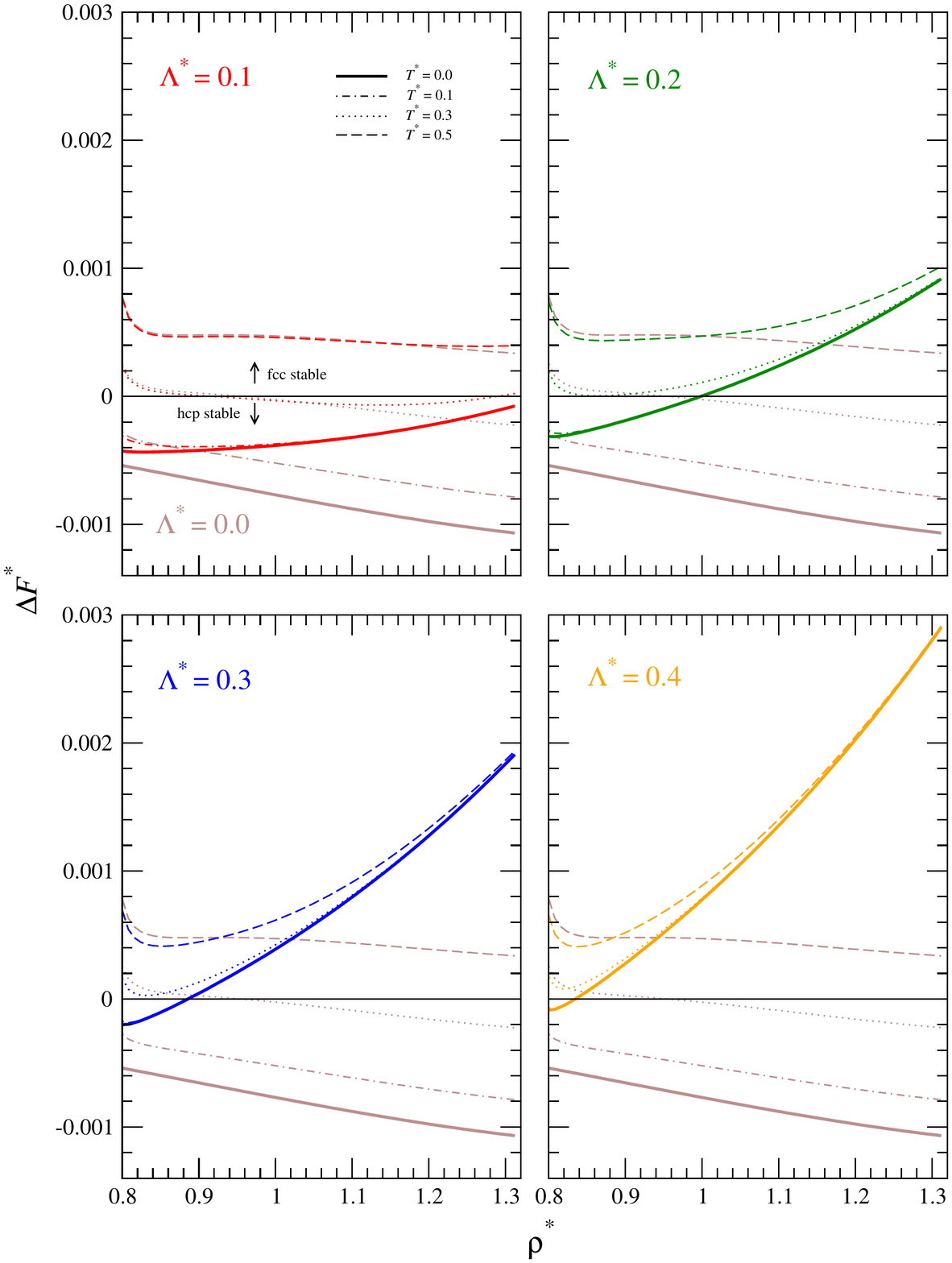}
    \caption{hcp-fcc free energy difference $\Delta F$ vs. $\rho$ for the LJ solid at $r_c^*=10$ obtained from QHLD, for various $T^*$ and $\Lambda^*$.}
    \label{fig:harmonic_DeltaF_vs_rho}
\end{figure}

\begin{figure}[!htbp]
    \centering
    \includegraphics[width=0.7\columnwidth]{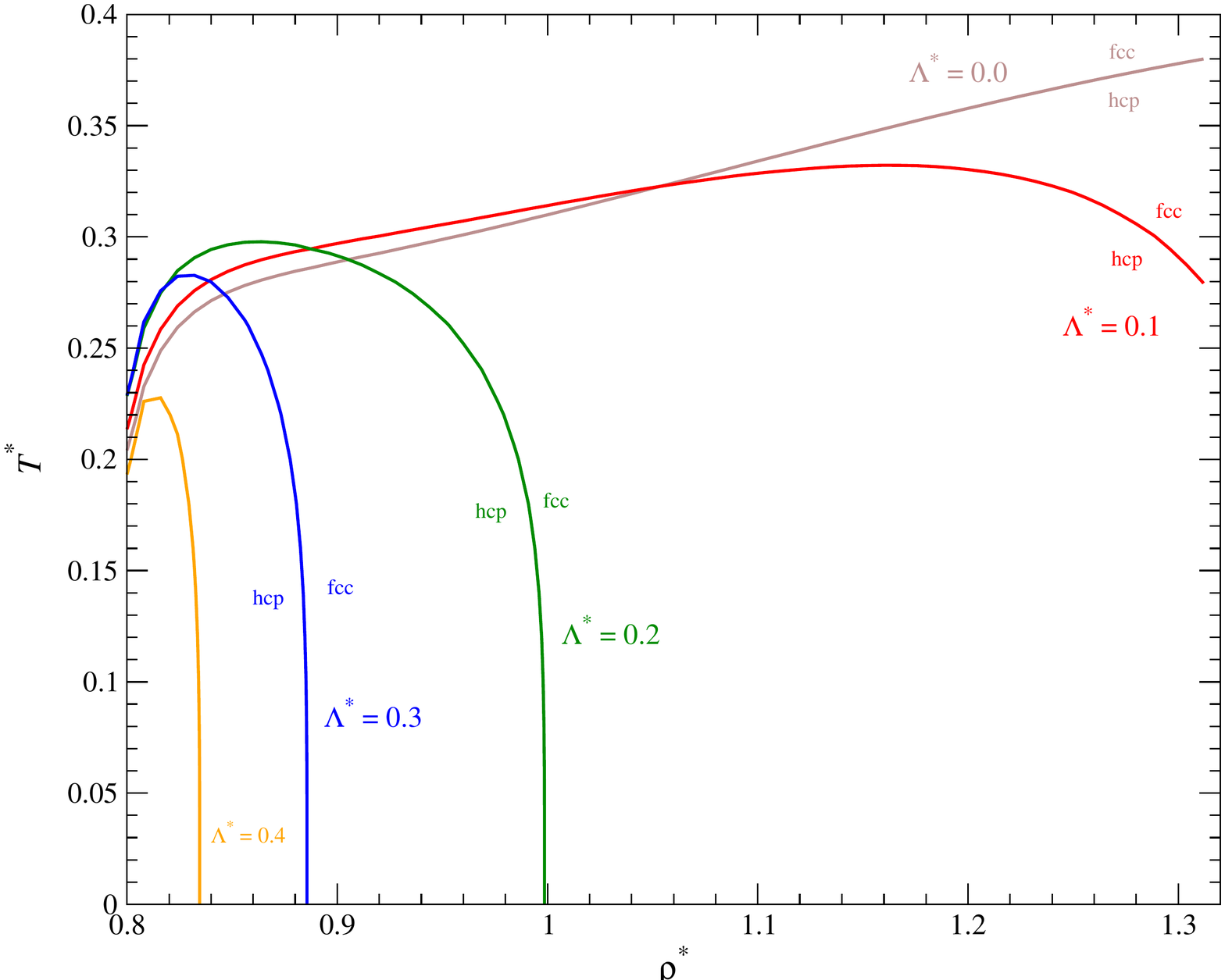}
    \caption{Phase diagram of the LJ solid in the $\rho^*$-$T^*$ plane, for various $\Lambda^*$, determined using QHLD and $r_c^*=10$.}
    \label{fig:harmonic_phase_diagram_NVT}
\end{figure}

We now consider the phase behaviour for $r_c^*=10$ over the density range $\rho^*=0.8$ to 1.3.
$\Delta F^*$ vs. $\rho^*$ is shown in Figure~\ref{fig:harmonic_DeltaF_vs_rho} for various $\Lambda^*$. 
The figure shows that increasing quantumness generally has the effect of stabilising fcc: as $\Lambda^*$ is increased, $\Delta F^*$ also increases. 
The effect on the phase diagram is shown in Figure~\ref{fig:harmonic_phase_diagram_NVT}, which shows the hcp-fcc phase boundaries for various $\Lambda^*$ in the $\rho^*$--$T^*$ plane. 
Note that increasing quantumness has the effect of reducing the size of the hcp region; the hcp region of the phase diagram is "compressed" towards $\rho^*=0.8$ as $\Lambda^*$ is increased. 

This behaviour is largely due to the zero-point energy. 
Recall that for $r_c^*=10$ we found for $\rho^*=1.07255$ that the zero-point energy favours fcc: $\Delta E^*_{ZP}>0$. 
The same is true at all other densities we considered. 
As the quantumness is increased, $\Delta E^*_{ZP}$ becomes a larger contribution to $\Delta F^*$, and therefore fcc becomes increasingly favoured. 
This is especially true at high densities, where, as shown in Figure~\ref{fig:harmonic_DeltaF_vs_rho}, $\Delta E^*_{ZP}$ is larger than at lower densities, leading to a more pronounced effect.

Of course, temperature also affects $\Delta F^*$ through the difference in the vibrational energies. 
At low pressures $\Delta E^*_{ZP}$ is smaller and temperature plays a larger role. 
At very low densities the phase behaviour is non-trivial: at $\rho^*=0.84$ as $\Lambda^*$ is increased there is an increase in the hcp-fcc transition temperature until $\Lambda^*=0.2$, followed by a decrease as $\Lambda^*$ is increased further.

\begin{figure}[!htbp]
    \centering
    \includegraphics[width=0.7\columnwidth]{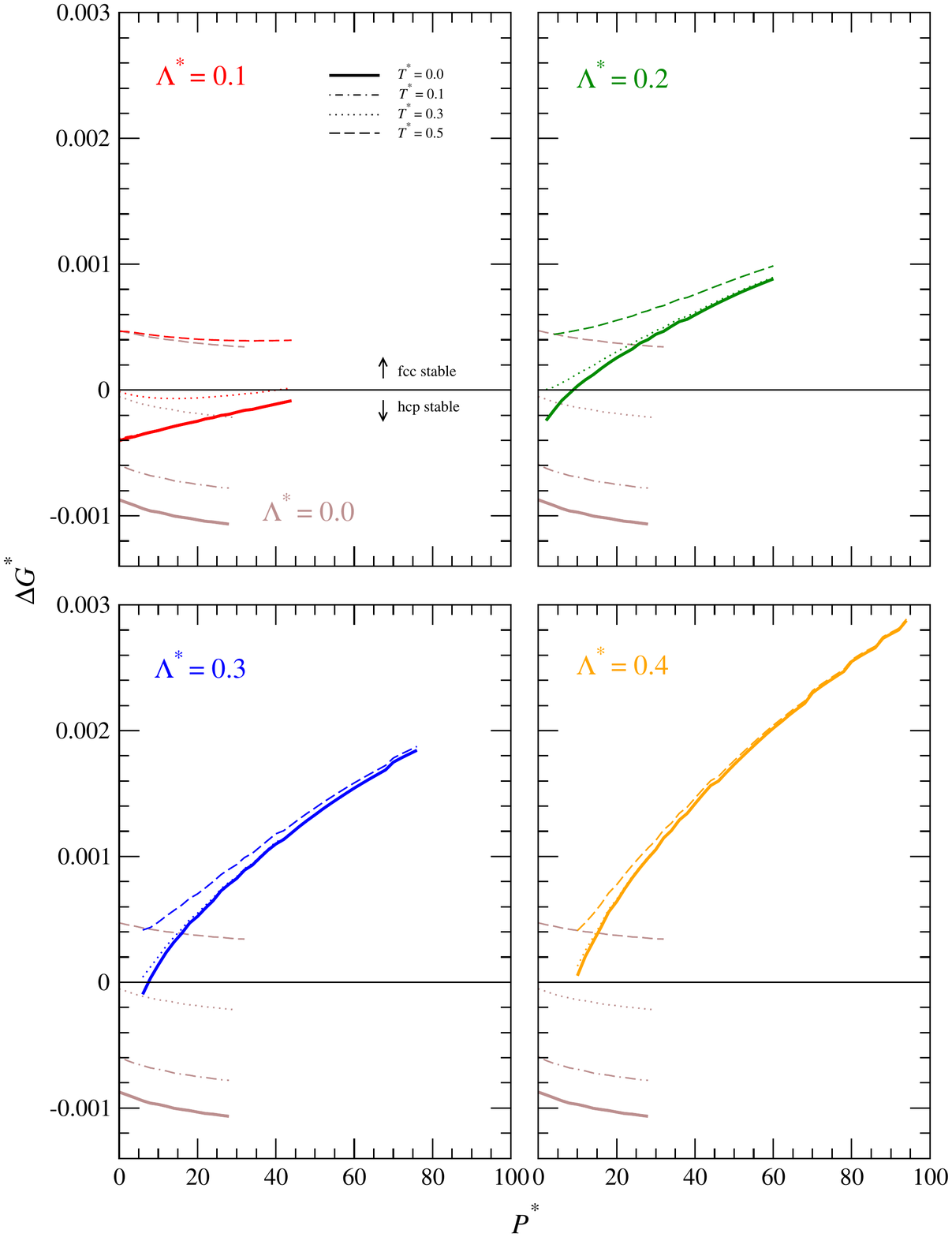}
    \caption{hcp-fcc free energy difference $\Delta G^*$ vs. $P^*$ for the LJ solid at $r_c^*=10$ obtained from QHLD, for various $T^*$ and $\Lambda^*$.}
    \label{fig:harmonic_DeltaG_vs_P}
\end{figure}

\begin{figure}[!htbp]
    \centering
    \includegraphics[width=0.7\columnwidth]{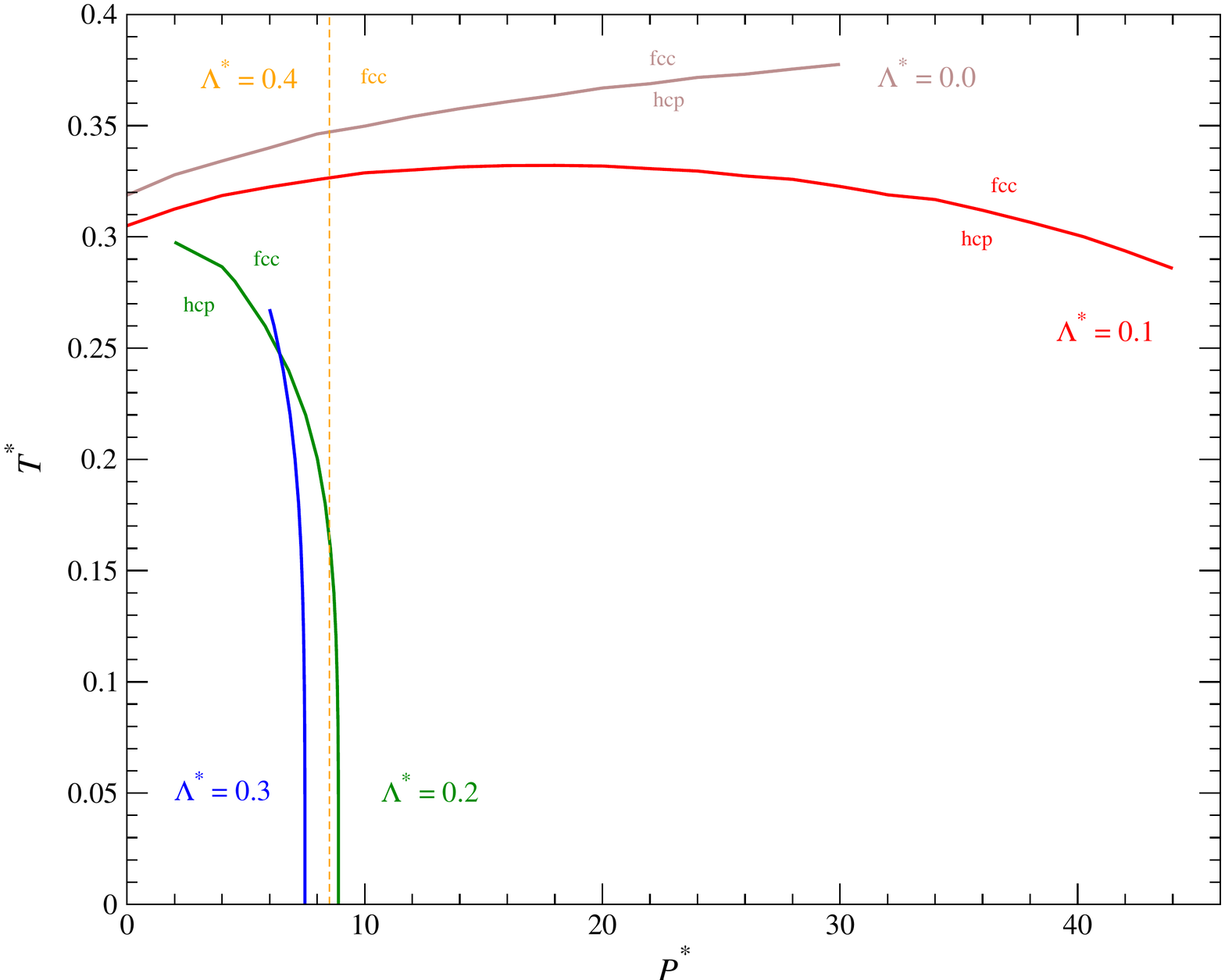}
    \caption{Phase diagram of the LJ solid in the $P^*$-$T^*$ plane, for various $\Lambda^*$, determined using QHLD and $r_c^*=10$. Solid lines correspond to phase boundaries for $\lambda=0.0$, 0.1, 0.2 and 0.3. No phase boundary exists for $\Lambda^*=0.4$; the fcc
    phase is stable for all realisable pressures. The lower bound for the realisable pressures for $\Lambda^*=0.4$ is indicated by a dashed line.}
    \label{fig:harmonic_phase_diagram_NPT}
\end{figure}

As can be seen from Figure~\ref{fig:harmonic_pressure}, much of the range covered in Figures~\ref{fig:harmonic_DeltaF_vs_rho} and \ref{fig:harmonic_phase_diagram_NVT} pertain to densities which correspond to negative pressures. 
$\Delta G^*$ vs. $P^*$ is shown in Figure~\ref{fig:harmonic_DeltaG_vs_P}, and the hcp-fcc phase boundaries in the $P^*$--$T^*$ plane are shown in Figure~\ref{fig:harmonic_phase_diagram_NPT}. 
Recall that neither crystal structure is mechanically stable at $P^*=0$ for $\Lambda^*\gtrsim 0.2$ in the quasi-harmonic approximation, which is why not all the curves in these figures extend to $P^*=0$. 
Figure~\ref{fig:harmonic_phase_diagram_NPT} reveals that as the quantumness is increased the location of the hcp-fcc transition is moved to lower pressures. 
Moreover, at $P^*=0$, increasing $\Lambda^*$ from 0 to 0.1 moves the $P^*=0$ transition
temperature to lower temperatures; the size of the region of hcp stability is reduced upon moving from $\Lambda^*=0$ to $\Lambda^*=0.1$. 
Interestingly, for $\Lambda=0.4$ there are no pressures at which hcp is thermodynamically stable. 
Hence no phase boundary exists in the quasi-harmonic approximation for $\Lambda^*=0.4$; the only stable phase is fcc.

\section{Discussion}\label{sec:discussion}

\subsection{Reconciling the two methods}
Our PIMD and QHLD calculations make qualitatively different predictions for the phase behaviour of the quantum LJ solid. 
At $P^* \approx 0$ the classical LJ solid exhibits a phase transition from the hcp phase (stable at $T^*=0$) to the fcc phase at $T^*\approx 0.3$. 
Our QHLD results suggest that this is also the case in the quantum solid for $\Lambda^* \lesssim 0.2$.
However, our PIMD results suggest that hcp is stable at $P^* \approx 0$ for all temperatures up to at least $T^*=0.5$.
Another clear discrepancy between the two methods arises with regards to the phase behaviour at low temperatures. 
QHLD implies that at $T^*\approx 0$ the hcp-fcc transition moves to lower densities as the quantumness is increased, going as low as $\rho^*=0.835$ at $\Lambda^*=0.4$. 
By contrast the PIMD transition density never drops below $\rho^*=1.0$ (see Figures~\ref{fig:quantumPhaseDiagrams} and \ref{fig:harmonic_phase_diagram_NVT}).
Finally, the hcp-fcc free energy differences obtained from PIMD are typically about an order of magnitude larger than those obtained from QHLD (see Figures~\ref{fig:quantumPhaseDiagrams} and \ref{fig:harmonic_phase_diagram_NVT}).

\begin{figure}[!htbp]
    \centering
    \includegraphics[width=0.7\columnwidth]{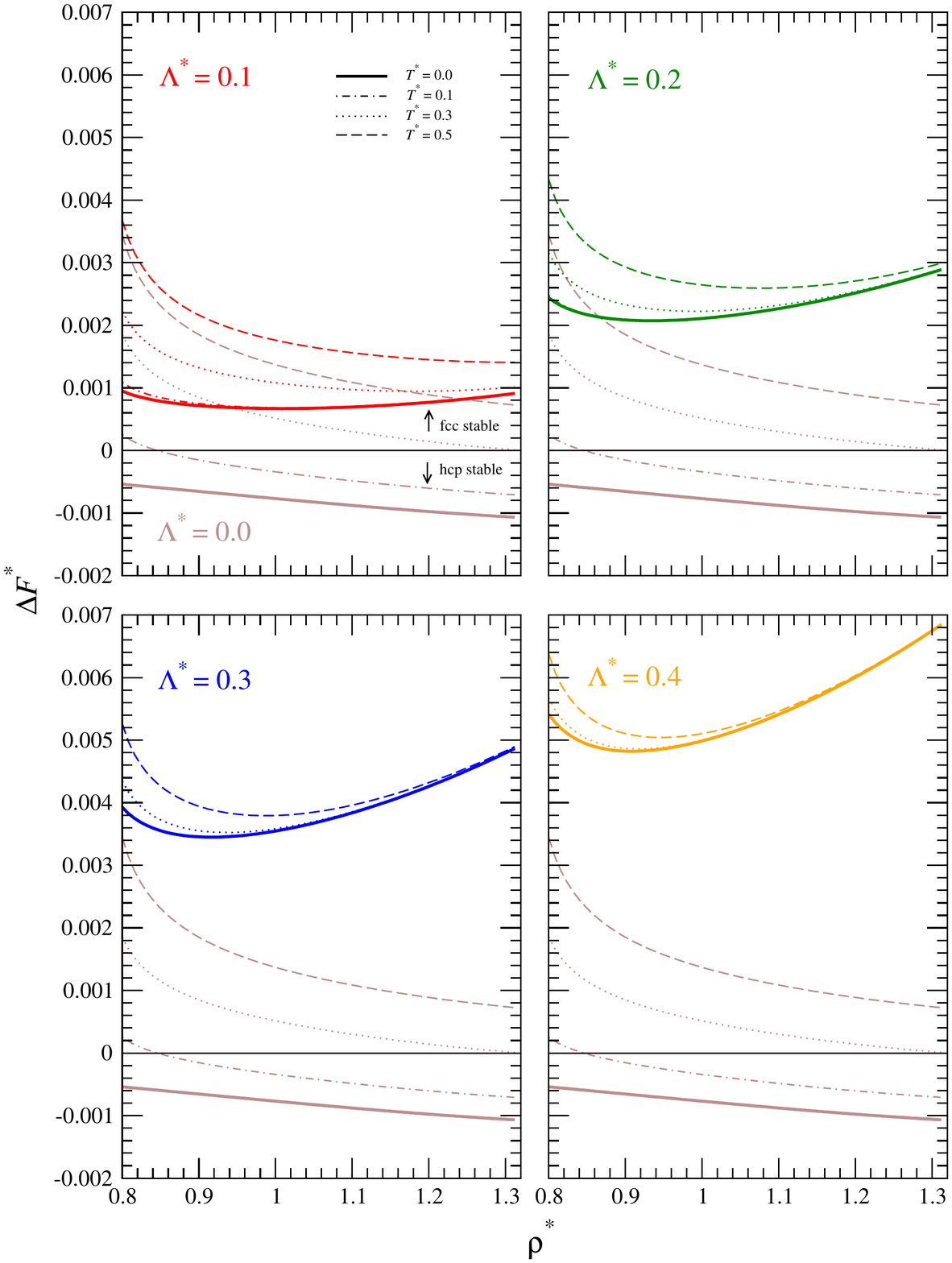}
    \caption{hcp-fcc free energy difference $\Delta F^*$ vs. $\rho^*$ for the LJ solid obtained from QHLD using the cutoff scheme described by Eqn.~\ref{eq:cutoffScaling}, for various $T^*$ and $\Lambda^*$.}
    \label{fig:harmonic_DeltaF_vs_rho_scaled_rc}
\end{figure}

There are a number of possible causes for the discrepancies between the two methods.
Firstly, the models used in the QHLD and PIMD calculations were different: the QHLD calculations employed $r_c^*=10$, while the PIMD calculations used a value of $r_c^*$ which scaled commensurately with the density (Eqn.~\ref{eq:cutoffScaling}).
To investigate this further, we repeated our QHLD calculations using the same cutoff scheme used in our PIMD calculations. 
However, we found that this \emph{worsened} the agreement between the two methods: using the scaled cutoff scheme makes the fcc \emph{more} stable in the quasi-harmonic approximation than it is for $r_c^*=10$. 
This can be seen by comparing Figures~ \ref{fig:harmonic_DeltaF_vs_rho} and \ref{fig:harmonic_DeltaF_vs_rho_scaled_rc}, the latter of which shows $\Delta F$ vs. $\rho^*$ using the scaled cutoff scheme.
Another key difference between our QHLD and PIMD calculations is that they pertain to different system sizes, namely $N=\infty$ and $N=256$ particles, respectively. 
It is well known that the quantitative details of the phase diagram for the classical LJ solid are sensitive to the system size, and surely the same is true for the quantum solid. 
However, we do not believe that finite size effects are the main cause of the discrepancies.
Rather, the main cause is the approximations which underpin the two methods.

\begin{figure}[!htbp]
	\includegraphics[width=0.7\columnwidth]{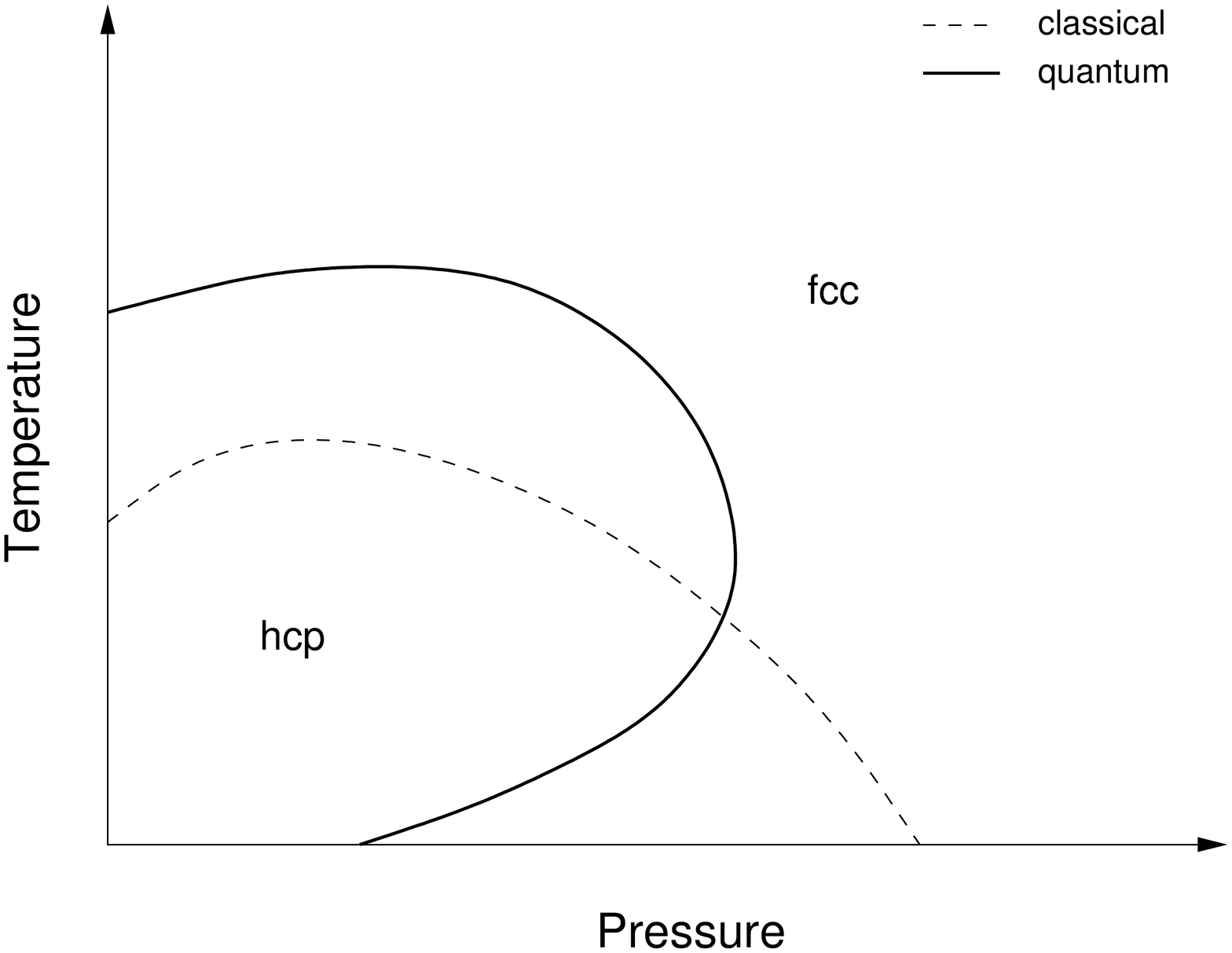}
	\caption{\label{fig:schematic_diagrmas} Schematic phase diagrams of the classical LJ solid, and a speculative quantum LJ solid where the hcp phase is re-entrant at moderate pressures.}
\end{figure}

To elaborate, QHLD is functionally exact in the zero-temperature limit but will break down at some finite temperature, and it is well known that this break down occurs very quickly in quantum molecular crystals\cite{Cazorla2017}.
PIMD, on the other hand, is more accurate than QHLD at finite temperature but calculations cannot be performed at temperatures near zero due to the prohibitive computational cost. 
The two methods therefore provide complimentary data: QHLD informs us about the zero-temperature limit, and PIMD speaks to the high-temperature limit. 
The true phase diagram of the quantum system may be a combination of the results from these two methods, where at low temperatures fcc is increasingly favoured as the quantumness is increased, and at high temperatures hcp is increasingly favoured.
This suggests a potential re-entrant fcc-hcp-fcc transition at low pressures for a highly-quantum LJ system, as illustrated schematically in Figure~\ref{fig:schematic_diagrmas}. 

\subsection{Origin of hcp stabilisation in PIMD}
\begin{figure}[!htbp]
	\includegraphics[width=0.6\columnwidth]{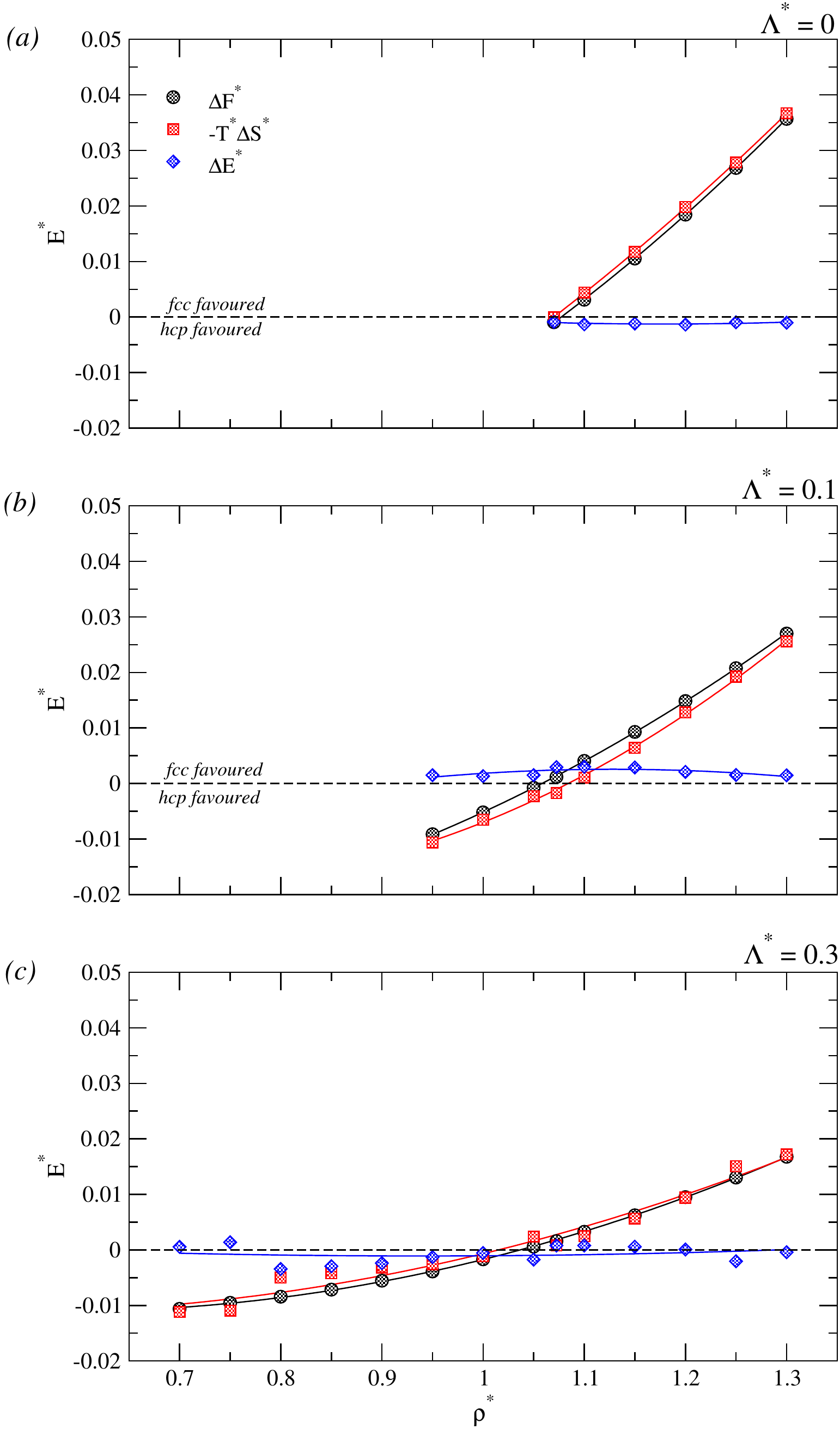}
	\caption{\label{fig:entropy-vs-energy} Contributions of entropy and internal energy to phase stabilization in the GSP-corrected model at $T^*=0.10$ for (a) $\Lambda^* = 0$ (classical system), (b) $\Lambda^* = 0.1$ and (c) $\Lambda^* = 0.3$. Negative values indicate stabilization of hcp and positive values indicate stabilization of fcc. Lines are quadratic fits to the data and are used to guide the eye.}
\end{figure}

The origin of the fcc stabilisation with increased quantumness in QHLD was shown to be due to its lower mean vibrational frequency and thus lower zero point energy. 
The origin of hcp stabilization in PIMD is less clear, due to the opaque nature of PIMD simulations. 
To investigate this further we partition the $\Delta F^*$ values from Figure~\ref{fig:quantumPhaseDiagrams} into contributions from internal energy and entropy. 
The average difference in internal energy $\Delta E^*$ can be extracted directly from the PIMD trajectories and so $-T^*\Delta S^*$ is easily accessible as:
\begin{equation}\label{eq:entropy}
-T^*\Delta S^*={\Delta F^*-\Delta E^*}
\end{equation}
The results are shown in Figure \ref{fig:entropy-vs-energy} for the cases of $\Lambda^* =$ 0, 0.1 and 0.3 at the reference temperature $T^* = 0.1$. 
Here we only consider the GSP-corrected model, as we do not wish to include fluctuations in the energy due to cutoff effects. 
The $\Delta F^*$ curve is reminiscent of the QHLD result in Figures \ref{fig:harmonic_DeltaF_vs_rho} and \ref{fig:harmonic_DeltaF_vs_rho_scaled_rc}, but larger in magnitude and shifted to more dramatically favour hcp at low densities. 
From these plots we can clearly see that $\Delta E^*$, which contains the zero point energy, is practically negligible and that the free energy difference between the two phases is due to the difference in entropy $-T^*\Delta S^*$. 
Therefore the stabilization of quantum hcp in PIMD must be due to anharmonic effects which are not accounted for in QHLD.

\subsection{Implications for noble gases}
\begin{figure}[!htbp]
	\includegraphics[width=0.6\columnwidth]{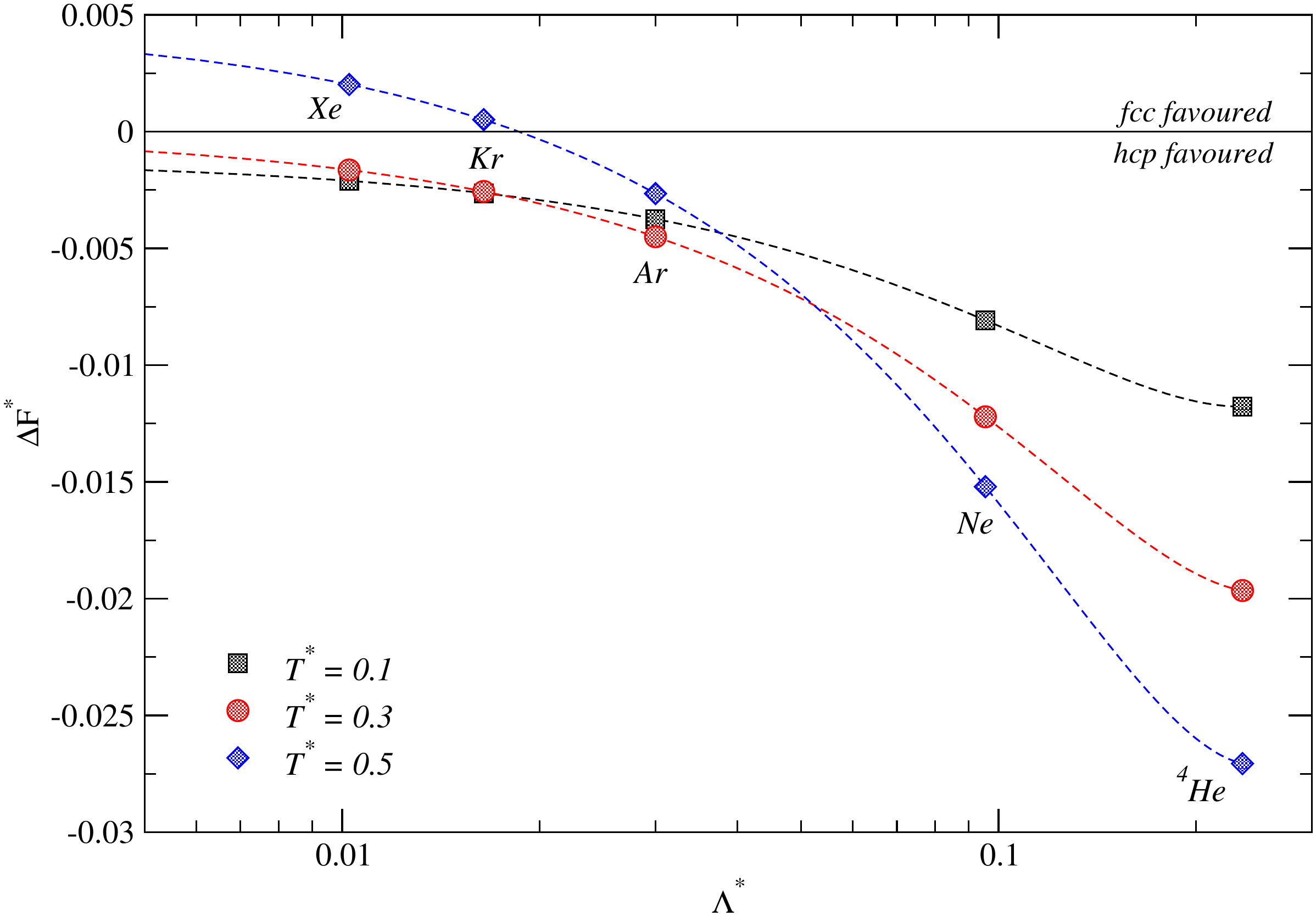}
	\caption{\label{fig:nobleGas} $\Lambda^*$ values for the noble gases (data points) superimposed on our quantum GSP-corrected $\Delta F^*$ results from Figure \ref{fig:quantumPhaseDiagrams} (dotted lines), represented here as zero-pressure isotherms at $T^* =$ 0.1, 0.3 and 0.5. A log scale is used to show detail near $\Lambda^*=0$. The $\sigma$ and $\varepsilon$ parameters needed to calculate $\Lambda^*$ were taken from ref. \onlinecite{Gordon1972} for helium and ref. \onlinecite{Rutkai2017} for Ne, Ar, Kr and Xe. }
\end{figure}

The LJ potential is often used to model interactions in noble gas solids and it is therefore interesting to consider our results in that context. 
Of the noble gases, only He has been observed in the hcp phase; all others adopt the fcc structure.\cite{Pollack1964,McMahon2012} 
Explanations for this phenomenon have been offered in the literature\cite{Cuthbert1958,Niebel1974,Borden1981} but a consensus has not been reached. 
In Figure \ref{fig:nobleGas} we compare the values of the de Boer parameter $\Lambda^*$ for the noble gases with our PIMD results for the GSP-corrected model. 

By and large these results agree with the experimental observations. 
The heavy noble gases (Ar, Kr and Xe) are clustered near the fcc phase, while He has a strong preference for hcp. 
The only discrepancy is Ne, which is shown to prefer hcp with this treatment but has an fcc structure in reality. 
However, the LJ potential is a somewhat simplistic treatment and a more realistic potential may show more accurate phase behaviour for Ne. 
Overall, our results show that the hcp structure of He is due to quantum effects, while the heavy noble gases prefer the classically-favoured fcc structure.  
Moreover, there are very large contributions of zero-point vibrations to the pressure, in the case of helium these are large enough to drive the density below the region in which the Lennard-Jones potential can stabilize a crystal structure.

\subsection{Prospect of other phases}
One key conclusion of our QHLD calculations was that $P=0$ was inaccessible for $\Lambda\gtrsim 0.3$ on account of the mechanical instability of the fcc and hcp crystals at very low pressures. 
Keeping in mind that QHLD is a perturbation method, the question remains as to whether any crystal structure \emph{is} stable for $\Lambda\gtrsim 0.3$ at such pressures.
In this work we have only considered the hcp and fcc phases, which are the only two phases known to be stable in the classical LJ solid. 
Of course, there is the prospect that phases other than hcp and fcc are stable at certain temperatures and low pressures for the quantum LJ solid. 

\section{Conclusion}\label{sec:conclusion}
Through a combination of PIMD simulations and lattice dynamics calculations, the inclusion of nuclear quantum effects in the LJ solid has been shown to stabilize the fcc phase at low temperature and the hcp phase at high temperature. 
In addition, the quantum effects on the phase behaviour of the quantum LJ solid in PIMD is relatively insensitive to the truncation scheme or cutoff length used, which makes it somewhat easier to draw definitive conclusions than their classical counterparts - despite the increased computational cost. 
We also offer an explanation for the experimentally observed phase behaviour of the noble gas solids. 
Helium, which is highly quantum, has a sufficiently large de Boer parameter that it falls squarely in the quantum-favoured hcp region. 
Thus the contrast between hcp helium and fcc structures for the other noble gases solids is purely due to quantum effects.

\section*{Acknowledgements}
This work is supported by the ERC grant HECATE (G.J.A. and H.W.) and the Engineering and Physical Sciences Research Council grant number EP/P007821/1 (T.L.U.). Computing resources were provided by the University of Edinburgh. This research made use of the Balena High Performance Computing (HPC) Service at the University of Bath.

\section{Data Availability}
The data that support the findings of this study are openly available from Edinburgh DataShare\cite{dataset} at https://doi.org/10.7488/ds/2846.

%

\end{document}